\documentclass[11pt,a4paper]{article}
\pdfoutput=1
\usepackage{jheppub}

\usepackage[top=4.5cm, bottom=0.cm, outer=-0.25cm, inner=4.75cm]{geometry}
\usepackage{placeins}
\usepackage{latexsym}
\usepackage[bbgreekl]{mathbbol}
\usepackage{mathtools}
\usepackage{amsmath,amssymb,amsthm,amsbsy,amsfonts,mathrsfs}
\usepackage{hyperref}
\usepackage{multirow}
\usepackage{slashed}
\usepackage{float}
\usepackage{bigints}
\usepackage{esint}
\usepackage{epsfig}
\usepackage{lscape}
\usepackage{titlesec}
\usepackage{enumitem}   

\usepackage{graphicx}

\usepackage{caption}
\usepackage{subcaption}
\usepackage[utf8]{inputenc}
\usepackage{nicefrac}
\usepackage{braket}
\usepackage{rotating}
\usepackage{afterpage}
\usepackage{bbm}
\usepackage{color,xcolor}
\usepackage{cancel}
\usepackage{stackrel}
\usepackage{standalone}
\usepackage[framemethod=tikz]{mdframed} 
\usepackage{empheq}
\usepackage[many]{tcolorbox}
\usepackage{simplewick}
\usepackage{pgf}
\usepackage{tikz}
\usetikzlibrary{shapes.geometric}
\usetikzlibrary{decorations.markings}
\usetikzlibrary{positioning}
\usetikzlibrary{fit}
\usepackage{tkz-berge}
\usepackage{rubikcube}
\usepackage{MnSymbol}
\usepackage{makecell}
\usepackage{float}

\usetikzlibrary{calc,patterns,angles,quotes}
\usetikzlibrary{decorations.pathmorphing}
\usetikzlibrary{arrows.meta}
\usetikzlibrary{decorations.markings}

\definecolor{darkpastelgreen}{rgb}{0.01, 0.75, 0.24 }
\definecolor{hooker\'sgreen}{rgb}{0.0, 0.44, 0.0}
\definecolor{indiagreen}{rgb}{0.07, 0.53, 0.03}
\definecolor{islamicgreen}{rgb}{0.0, 0.56, 0.0}
\usepackage{tikz-3dplot}

\makeatother
\newmdenv[%
middlelinecolor=blue!20!,
middlelinewidth=1pt,
backgroundcolor=blue!10!,
roundcorner=10pt
]{identity}

\newmdenv[%
middlelinecolor=red!20!white,
middlelinewidth=1pt,
backgroundcolor=red!10!white,
roundcorner=10pt,
subtitlebelowline=true,
frametitle={Calculation},
frametitlefont={\normalfont\bfseries\sffamily\color{red!40!white}},
]{calculation}

\include{macros}

\def\bbx#1\ebx{\begin{empheq}[box={\tcbhighmath[colframe=blue!20!white,colback=blue!10!white]}]{align} #1 \end{empheq}}

\title{Flat holography and celestial shockwaves}

\author{Zezhuang Hao}
\author{and Marika Taylor}

\affiliation{STAG Research Centre, Highfield, University of Southampton, SO17 1BJ Southampton, UK}  
\affiliation{School of Mathematical Sciences, Highfield, University of Southampton, SO17 1BJ Southampton, UK}
\emailAdd{Z.Hao@soton.ac.uk}
\emailAdd{M.M.Taylor@soton.ac.uk}

\abstract{In this paper we systematically develop the flat/CFT holographic dictionary, building on AdS/CFT holography. After analysing the behaviour of scalar field modes on hyperbolic slices of Minkowski and performing the holographic renormalisation for the associated onshell action, we obtain a holography dictionary between the bulk theory and the corresponding dual theory on the celestial sphere. We propose that a single scalar field in the bulk is dual to two series of operators on the celestial sphere; the scaling dimension of these operators takes values on the principal series. The real time features of the bulk theory, such as the dynamical and the casual structure, are encoded in the construction of correlation functions on the boundary via the coefficients of the bulk modes. Moreover, we will see that the two series of operators can be interpreted as ingoing and outgoing waves in the bulk. We illustrate our dictionary with the example of a single shock wave. Our results lay foundations for further computation within the flat/celestial CFT correspondence. }

\begin{document}  
\maketitle

%\newpage
%\tableofcontents
%\addtocontents{toc}{\protect\setcounter{tocdepth}{3}}
%\renewcommand{\theequation}{\arabic{section}.\arabic{equation}}

%%%%%%%%%%%%%%%%%%%%%%%%%%%%%%%%%%%%%%%%%%%%%%%%%%%%%%%%%%%%%%%%%%%%%%%%%
%%%%%%%%%%%%%%%%%%%%%%%%%%%%%%%%%%%%%%%%%%%%%%%%%%%%%%%%%%%%%%%%%%%%%%%%%
\section{Introduction}

\qquad General relativity and quantum mechanics have been brought up for a century and they are believed to be the most fundamental rules which respectively govern the large scale structure of the universe and the microscopic interactions between elementary particles even though they are not compatible with each other. After their great success in predicting  the observation from the lab, physicists spent long time looking for a unified theory of quantum and gravity, e.g. semiclassical field theory, supergravity, string theory. The quantum gravity theory that would please everyone has not been figured out yet while, during the extensive study of various proposed models, another fundamental principle which relates the dimension of spacetime, quantum and gravity effects has been found and caused a great attention in this century, so called holography principle.

The idea of projecting the physical world to a lower dimension one living on the boundary exists for long while it has not been formally discussed in physics literature until the work \cite{tHooft:1993dmi,Susskind:1994vu}, initiated by the study of the black hole entropy \cite{Christodoulou:1970wf,Penrose:1971uk,Bekenstein:1973ur,Hawking:1975vcx} which tells us that the entropy of the black hole is proportional to its horizon area. 

Based on such observation, one can further conclude that the degrees of freedom or information for a given system is bounded by its area of boundary rather the volume, which makes it possible to encode all the bulk information into the proposed boundary system. Such correspondence concerning the assignment of degrees of freedom is then developed to the duality between the subregion of the bulk and the boundary, conjectured to be characterised by the Ryu-Takayanagi surface \cite{Ryu:2006bv}. After proper assumptions, the conjecture was proofed in \cite{Lewkowycz:2013nqa,Faulkner:2013ana} and then, taking the quantum effect into consideration, the concept of RT surface is generalised to the surface named quantum extremal surface \cite{Engelhardt:2014gca}. 

Here we will not follow the stream of the discussion of entropy thus it turns out that the structure of holography is more than the projection of degrees of freedom after the first concrete realization of the holography principle discovered by \cite{Maldacena:1997re}, called AdS/CFT correspondence. In that work, Maldacena pointed out that type IIB string theory on the $AdS_5\times S^5$ background is dual to the super-Yang-Mills theory in 3+1 spacetime dimensions by studying the decoupling limit of the stack of $D_3$ branes in string theory and its corresponding low energy supergravity solution, which implies that, in additional to the reduction of dimension, the theory of quantum and gravity could also be relevant when comparing the theory in the AdS bulk with its boundary CFT correspondence. Such relation works exactly like the relativity between space and time in gravity or the relativity between the particle and wave in quantum.

However in practice, due to the lack of knowledge for the quantum gravity theory and the difficulty of studying strongly coupled gauge theory at low energy, one can first choose to investigate the AdS/CFT correspondence at the 't Hooft large $N$ limit, under which the gauge theory will be simplified since the contribution from planar diagrams will become dominant if the number of colors $N$ is large when keeping $\lambda=g^2_{YM}N$ constant \cite{tHooft:1973alw}. From the bulk side, we see that the string theory will become classical by comparing the map between parameters $g_s\sim g_{YM}^2$ and  $\alpha\sim 1/\sqrt{g_sN}$. Moreover, by taking large value of $\lambda$, the bulk theory will be weakly coupled thus the AdS/CFT correspondence becomes a weak/strong duality. In such case, the bulk theory is described by the semiclassical field theory and one can write down the effective action, decomposing the field at the boundary, and then map the data from asymptotic AdS infinity to the boundary CFT named AdS/CFT dictionary. 

In the literature, there are mainly two ways to construct the AdS/CFT dictionary \cite{Witten:1998qj,Gubser:1998bc}. One starts from the effective field theory on $AdS_5\times S^5$ background while the other starts from $AdS_5$ thus they are called top -down and bottom-up approaches to AdS/CFT, respectively. At first sight, the bottom-up approach looks easier if one just consider the fields on the $AdS_5$ background but the supersymmetric information is lost since  the ignorance of the Kaluza-Klein fields on the $S^5$ sphere, e.g. we would obtain non-zero vacuum energy. Such issue is rescued in the work \cite{Skenderis:2006di,Skenderis:2006uy} by Kostas and Marika. They developed a KK reduction map which reduces all the fields in 10d to 5d in a gauge invariant way therefore concludes that the top-down and bottom-up approaches could be equivalent provided that proper reduction procedure is applied. For this article, we will adopt the bottom-up approach and ignore the KK fields on the internal space. In this case, the duality is clarified by the dictionary proposed by Witten 
\begin{equation}
    {\rm exp}\;\Big(- S_{{\rm AdS}_{d+1}}(\Phi)\Big)_{\Phi\sim \phi_0}=\Big\langle\; {\rm exp}\;-\int_{S^{d}} \phi_0\;\mathcal{O}\;\Big\rangle_{CFT},\label{ads/cft}
\end{equation}
in which $S_{{\rm AdS_{d+1}}}(\Phi)$ is the action of the semi-classical theory in the bulk with scalar fields characterised by the asymptotic behaviour $\Phi\sim\rho^{-d+\Delta}\phi_0$ at large AdS radius $\rho$. From the right hand side, we can see that $\phi_0$ is dual to the source in the CFT theory and it is coupled to the operator $\mathcal{O}$. The scale dimension $\Delta$ of the operator and the mass $M$ of the particle in the bulk preserve the relation $\Delta(\Delta-d)=M^2$, which is obtained by solving the equation of motion of $\Phi$.

During last two decades, the AdS/CFT correspondence has caused a great interest among physicists and the general correspondence principle itself or the dictionary \eqref{ads/cft} has been verified in a quite amount of work, by constructing proper models and comparing calculation results between two sides. For example one can see the work on HS/CFT correspondence \cite{Sezgin:2002rt,Klebanov:2002ja,Sezgin:2003pt,Giombi:2009wh,Aharony:2020omh} and the work for ${\rm AdS_3/ CFT_2}$ correspondence \cite{Gaberdiel:2010pz,Eberhardt:2018ouy,Eberhardt:2019ywk}. However, although the AdS/CFT correspondence has passed through all theoretical tests, it is still far from being tested in the lab or producing any non-trivial predictions of the physical world. One of the main issue here is that, on the bulk side of the duality, the geometry background is AdS while the measurement of cosmology tells us that the geometry of our universe is with small positive curvature near to the flat spacetime.

There are developments on the dS/CFT correspondence \cite{Strominger:2001pn,Witten:2001kn} by matching the generators of the isometry group of de Sitter and the boundary conformal symmetry group while it is still not clear how to decompose the boundary data of fields on dS and map them to the boundary CFT therefore construct the dS/CFT dictionary. As for the flat case, following the idea from \cite{Polchinski:1999ry,Giddings:1999jq,deBoer:2003vf}, people have proposed that scattering amplitudes in Minkowski are dual to correlation functions of celestial CFT living on the celestial sphere. Such proposal has been developed dramatically in recent years (One can see the lecture notes \cite{Strominger:2017zoo,raclariu2021lectures,Pasterski:2021rjz} and references therein.) but the approach goes beyond from the standard treatment of AdS/CFT. 

Actually, there have been attempts to address the problems for developing a flat version of holographic principle dated 
 even back to the birth of AdS/CFT in the talk given by Witten \cite{98talk}. During that talk, he discussed various obstacle to writing down the flat/CFT dictionary. Conceptually, if one assumes that both of the quantum gravity theory and scattering amplitudes are dual to the CFTs on the boundary, then it will be hard to understand that why the quantum gravity theory should be equivalent to its own scattering amplitudes. From the technical point of view, the complexity for the geometric structure and the behaviour of fields at the two null boundaries of Minkowski space make it hard to write down the boundary correlation functions or to study the distribution of the degrees of freedom. At the end, he proposed that if the flat theory is dual to the  structure $X$ on the boundary, then $X$ should be more complicated than CFT. The complicated nature of the structure $X$ can also be seen from the study of symmetries of the asymptotic flat spacetime. Not like the AdS case, the isometry group for asymptotically flat space will reduce to the infinite BMS group \cite{Bondi:1962px,Sachs:1962wk,Sachs:1962zza} rather than the Poincar$\acute{{\rm e}}$ group. Globally, the BMS group is generated by the supertranslations and superrotations in which supertranslations behaves like 1d translation while superrotations are characterised by $SL(2,\mathbb{C})$. After fifty years of study of BMS group, people realised that the superroations could be locally generalised to the Virasoro even with central extension \cite{Barnich:2009se,Barnich:2010eb,Barnich:2011mi}, which brings hope to construct the duality between the flat theory and the 2d CFT \cite{Belavin:1984vu}.

 Based on another observation that the supertranslation Ward identity is equivalent to the Weinberg's soft graviton theorem on the celestial sphere \cite{Strominger:2013jfa,He:2014laa} when studying the symmetry of the graviton scattering amplitudes, Strominger with his collaborators then conjectured the duality between scattering amplitudes and celestial CFT so called celestial holography. After that, many properties concerning the celestial CFT have been explored e.g. they have constructed the dual celestial stress tensor \cite{He:2015zea} and the corresponding celestial CFT OPE coefficients are also discussed \cite{Pate:2019lpp,Guevara:2021abz,Strominger:2021lvk}. Although the celestial CFT exhibits a rich structure for people to study the scattering amplitudes, it is different from the standard 2d CFT and there are some other issues which cause a great confusion. For example, people could not understand the reason why a real time flat theory should be dual to an Euclidean theory on the sphere and it is still hard to say if the celestial CFT is unitary or not since the scale dimension living on the principle series are complex. There are later developments which claim that the $4d$ scattering amplitudes should be dual to the $3d$ Carrollian CFT \cite{donnay2022carrollian,Bagchi:2022emh} thus the BMS symmetry is manifested and signatures from both sides will fit.  Here we are not going to follow their approach while one can see that some of the problems will be clear once the structure $X$ is specified and a proper flat/CFT dictionary is given. 

 The main goal of this article is to develop the AdS/CFT correspondence into the flat/CFT correspondence thus bring the holography principle and especially all the work on AdS/CFT to the measurable level. More precisely, we will finally construct the dictionary between flat spacetime and the CFT on the boundary which works the same as  \eqref{ads/cft}.
 
Before going into the flat/CFT dictionary, we first introduce the other principle used many times in this article, which is the completeness relation of mode expansion for a generic physical field which tells us that a generic field configuration can be decomposed into given modes for the linear physical system and the information of the field is encoded in coefficients. They are determined by the boundary data. Physically, for example in quantum mechanics and quantum field theory, one always assumes that all the modes form a complete set of basis for the physical solution space while the rigorous mathematical structure has been studied so-called Strum-Liouville theory even though the boundary conditions are often hard to specify or to check in the physical situation.

In physics the mode expansion is also called superposition principle and has been used widely dated from the birth of quantum mechanics. Here we will reconsider the mode expansion and find it is not as obvious as people thought it would be although it has been taken for granted for long. Taking the story of quantum field theory for example, the traditional mode is the plane wave mode $\Phi_K=e^{iK\cdot X}$ and all on-shell modes satisfying $-K^2=M^2$ form a complete basis for the field describing particles of mass $M$. After the quantisation, coefficients for the plane wave are promoted to be creation and annihilation operators. Recently, except for plane waves, people have constructed a new kind of basis so called conformal basis $\Phi_\Delta$ \cite{Pasterski:2016qvg,Pasterski:2017kqt} to highlight the symmetry of Lorentz group $SO(1,3)$ and the unitarity of the representation of Lorentz group requires that $\Delta$ should lie on the principal series thus one assumes that all states on the principal series form a complete basis. In addition to the conformal basis, for this article, we are going to introduce another kind of mode based on the foliation of the Minkowski as
\begin{equation}
    -(X^0)^2+(X^1)^2+(X^2)^2+(X^3)^2=-\tau^2,\label{AdSslice}
\end{equation}
in which one can treat it as the embedding of the AdS hyperboloid with radius $\tau\geq 0$. Given such foliation, one can further choose $\tau$ together with the coordinate on AdS surface as Minkowski space coordinates therefore recast the equation of motion for Minkowski into the AdS hyperboloid. Then, according to the superposition principle, one can claim that a generic field can be decomposed into modes $\Phi_k$ with effective mass $k$ on the AdS surface. Since here the equation on AdS is not physical, $k$ could take all the value in the complex plane and till now it is not clear how to determine which of them will form the necessary complete basis. From the boundary point of view, it leaves the range of scale dimension $\Delta_k$ undetermined since we have the dictionary
\begin{equation}
    \Delta_k(\Delta_k-2)=k^2.
\end{equation}
We will use Klein-Gordon equation as an example to illustrate how the mode expansion works  
 in the context of AdS slicing \eqref{AdSslice} and discuss various possible choices of $k$ in the section \ref{Mode} by exploring the physical meaning and the stability of the on-shell modes.

After a careful study of the mode analysis, then one will be able to decompose the bulk action $S$ for Minkowski space into $k$-mode components $S(k)$ like what has been done for on-shell fields. To construct the flat/CFT dictionary like \eqref{ads/cft}, a technical issue ahead is that the on-shell action $S^{{\rm onshell}}$ is infinite due to the integral over the infinite spacetime volume and one needs to perform the renormalisation on $S^{{\rm onshell
}}$ in order to make the action finite, denoted as $S^{{\rm ren}}$ or equivalently $S^{{\rm ren}}(k)$. Such problem was addressed in the work \cite{Witten:1998qj} then has been fully discussed by following work \cite{Henningson:1998gx,Balasubramanian:1999re,deHaro:2000vlm,skenderis2002lecture,Papadimitriou:2004ap}. The developed systematic procedure is so called holographic renormalisation. The basis idea of holographic renormalisation is that one should treat the infinite part of the action in the bulk as IR divergences and introduce local counterterms $S^{{\rm ct}}$ to cancel the divergence, i.e. $S^{{\rm ren}}=S^{{\rm onshell}}+S^{{\rm ct}}$. Such IR divergences in the bulk are dual to the UV divergences of the boundary QFT through the UV/IR connection \cite{Susskind:1998dq}. The UV divergence in the bulk is dual to the IR divergence of the boundary QFT while it should be absent when working in the full context of the holography principle since the bulk quantum gravity theory is UV finite. As for the low energy effective description of the bulk theory, the UV divergence will appear and contribute to anomalous dimensions of CFT operators from boundary point of view. We will not discuss them in this article and  one can see the straightforward treatment of UV divergences from the bulk side in the recent work \cite{Banados:2022nhj}.  In the section \ref{Dictionary}, we will first decompose the field into AdS modes then apply the holographic renormalisaton procedure on each single AdS surface thus complete the holographic renormalisation for flat spacetime.

Given the flat holography renormalisation, one then obtains the dictionary between the effective theory on Minkowski and the CFT living on the boundary sphere. The CFT context can be read off from the renormalised action $S^{{\rm ren}}$ and it turns out that a single bulk scalar field is dual to two series of CFT operators on the sphere with scale dimension living on the principal series. \footnote{The idea that a single field is dual to infinite operators on principal series was pointed out by de Boer and Solodukhin \cite{deBoer:2003vf} then it is widely used in Celestial holography to study the amplitudes in the bulk by matching the symmetries between the scattering amplitudes and the celestial operators. Here we should emphasis that, in this article, there are two distinguished series of CFT operators living on the same sphere in order to fully describe a scalar field in a single Milne wedge. There is so called celestial holography dictionary \cite{Strominger:2017zoo,raclariu2021lectures,Pasterski:2021rjz} which directly gives us the relation between bulk modes and celestial CFT operators. But to address the dynamics of the theory, following the development of celestial holography, one further needs to specify the relation between collinear limits of the amplitudes and the OPE coefficients of the celestial CFT based on the spirit of bootstrap.  Strictly speaking, we should note that the definition of the operators in celestial holography and the operators in this article are different. But they have similar physical meaning which can be used to describe the in and out going modes and they should be equivalent. We will briefly discuss the difference and connection in section \ref{Shock}.} Here we construct the dictionary in a precise way thus one should be able to give a full description of the bulk quantum gravity theory by studying the boundary field theory or vice versa. From the dictionary we can use the bulk action to produce all the boundary correlation functions thus one can see the relation between the dynamics of the bulk theory and the coupling of the boundary operators, i.e. the coupling way is determined by the dynamical and causal structure of the bulk theory. Moreover, corresponding two-point and three-point correlation functions on the celestial sphere are also studied in the context of $\Phi^3$ and $\Phi^4$ interaction and they are represented as double-disk diagrams.

Later in section \ref{Shock}, we will see that we can decompose a massless field into in and out going shock waves and each of the shock wave is dual to one series of operators on the celestial sphere. One-point and two-point functions on the celestial sphere dual to the shock wave are also derived. For spherical shock waves, the two-point function will become trivial at leading order while the subleading term relies on the study of backreaction of the metric and the broken of the spherical symmetry in the perturbative sense. As a simple model, we take the spherical shock wave as an example to perform the mode analysis procedure introduced in section \ref{Mode} and the corresponding coefficients are determined.  Furthermore, we find that the full information of Minkowski could be stored in a pair of AdS hyperboloid, which forms a new kind of Cauchy surface.

\section{Mode Analysis on Minkowski}\label{Mode}

\qquad In this section we consider solutions of the scalar field equation on Minkowski space and discuss how these can be used to construct a basis for scalar fields satisfying 
the given boundary conditions. We will begin our discussions with the familiar analysis within Minkowski coordinates before moving to Anti-de Sitter and de Sitter slicings. 

Let us begin with the massive scalar equation
\begin{equation}
    \left( \frac{\partial}{\partial X^\mu}\frac{\partial}{\partial X_\mu}-M^2\right)\Phi_M(X)=0,\label{M}
\end{equation}
in which $M$ is the mass of the scalar field $\Phi_M$ and $X^\mu = \{ X^0, X^i \}$ are coordinates of the Minkowski space $\mathbb{R}^{1,3}$ with signature $(-,+,+,+)$. 

One can immediately write down a basis for solutions for this equation
\begin{equation}
    f_K(X)=e^{i K \cdot X}
\end{equation}
in which $K^2+M^2=0$ and $K^\mu$ is understood as the momentum of the particle. The restriction of $K^{\mu}$ to be real follows from imposing boundedness of the field as either $X^0$ or $X^i$ approach infinity; this is implicitly assumed in most analyses.  A generic scalar field $\Phi$ satisfying these boundary conditions can then be expressed as usual as
\begin{equation}
\Phi (X) = \int d^4 K \; \Phi (K) f_K(X)
\end{equation}
For an onshell field of mass $M$ the field in momentum space is such that $\Phi_M(K) \propto \delta (K^2 + M^2)$. 

The approach above intrinsically respects relativistic covariance. In some contexts one works with bases that partially break this covariance, particularly by separating space and time. The basis above can trivially be rewritten as
\begin{equation} 
f_{w,k} (X^0,X^i) = e^{i w X^0} e^{i k_i X^i}, 
\end{equation}
where the mass-shell condition is $\omega^2 = k^i k_i + m^2$. Again this basis can be used to express any scalar field with the same boundary conditions. One can also use a mixed representation to express a field e.g. 
\begin{equation}
\Phi(X^0,X^i) = \int d^3 k \; \Phi(X^0,k^i) e^{i k_i X^i}
\end{equation}
although the onshell condition for the mixed representation field is then a differential equation rather than algebraic i.e. 
\begin{equation}
\partial_{X^0}^2 \Phi(X^0,k^i) = (k^2 + M^2) \Phi(X^0,k^i). 
\end{equation}

\begin{figure}[h!]
    \centering
    \begin{tikzpicture}
\draw[thick](0,3)--(3,0);
\draw[thick](3,0)--(0,-3);
\draw[thick](-3,0)--(0,-3);
\draw[thick](-3,0)--(0,3);
\draw[thick](-1.5,-1.5)--(1.5,1.5);
\draw[thick](-1.5,1.5)--(1.5,-1.5);
\draw[thick,blue](-1.5,-1.5)..controls(0,-1)..(1.5,-1.5);
\draw (3.3,0) node{$i_0$};
\draw (0,3.3) node{$i^+$};
\draw (0,-3.3) node{$i^-$};
\draw [thick,blue](-1.5,1.5)..controls(0,1)..(1.5,1.5);
\draw [thick,blue](1.5,1.5)..controls(1,0)..(1.5,-1.5);
\draw (0,-1.5) node{$\mathcal{A}^-$};
\draw(0,1.5) node{$\mathcal{A}^+$};
\draw (1.5,0) node{$\mathcal{D}$};
\end{tikzpicture}    
    \caption{The Milne wedge $\mathcal{A}^\pm$ are sliced by AdS surfaces while the Rindler wedge $\mathcal{D}$ is foliated by dS surfaces.}
    \label{slicing}
\end{figure}
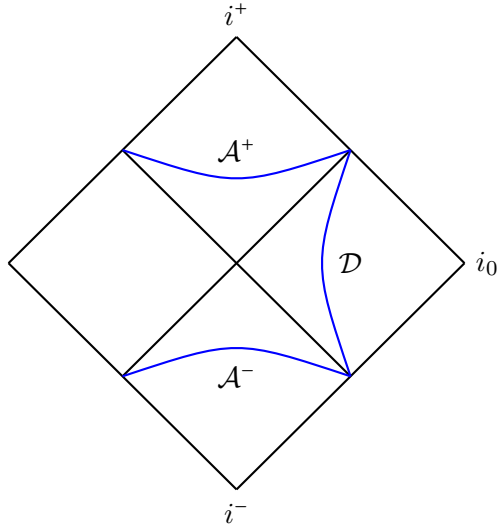
\subsection{Milne Slicing}

\qquad Following this review, we now consider solution of the scalar field equation using Anti-de Sitter and de Sitter slicing of Minkowski space. We illustrate these slicings in Figure~\ref{slicing}.  Region $\mathcal{A}^\pm$ are foliated by Euclidean Anti-de Sitter (hyperbolic) surfaces while region $\mathcal{D}$ is foliated by de Sitter surfaces. 
To describe the region $\mathcal{A}$ which is sliced by hyperboloids, we use Milne coordinates written as
\begin{equation}
    ds^2=G_{\mu\nu}dX^\mu dX^\nu=-d\tau^2+\tau^2\left(\frac{d\rho^2}{1+\rho^2}+2\rho^2 \gamma_{z\bar{z}}dzd\bar{z} \right),\label{milen}
\end{equation}
in which $\rho,\tau\in \mathbb{R}$. Here $z,\bar{z}$ are complex coordinates and the metric $\gamma_{z\bar{z}}$ is the standard metric on the sphere. $\tau$ is the radius of the AdS hyperboloid introduced in \eqref{AdSslice} and one only needs to take the positive part $\tau\geq 0$ to cover the single region $\mathcal{A}^+$. The Milne horizon is given by $\tau\rightarrow 0$ while the null infinity is the region where $\tau\rightarrow \infty$. In such coordinates, the scalar equation can be separated into two equations
\begin{eqnarray}
\left(\rho(\rho^2+1)\partial^2_\rho+(3\rho^2+2)\partial_\rho-k^2\rho-\frac{l(l+1)}{\rho}\right) \phi_{l}(\rho, k)&=&0,\label{effective}\\
    \left(-3\frac{\partial_\tau}{\tau}-\partial^2_\tau+\frac{\omega^2}{\tau^2}-M^2\right) \psi(\tau, \omega)&=&0,\label{ms}
\end{eqnarray}
where the first equation represents a particle of effective mass $k$ on the hyperboloid and the second equation depends only on the time $\tau$. Here $l$ labels the usual discrete eigenvalue of scalar spherical harmonics $Y^l_m(z,\bar{z})$. Accordingly the scalar basis can be expressed as
\begin{equation}
f_{\omega, k, l, m} (\tau, \rho, z, \bar{z}) = \psi(\tau, \omega) \phi_{l}(\rho, k)  Y^l_m(z,\bar{z})
\end{equation}
where the onshell condition requires $\omega = k$. As above, we will be interested in using this basis to represent fields with the same boundary conditions which are not necessarily onshell, hence we do not impose $\omega = k$ a priori. Now, based on the superposition principle, any scalar satisfying the boundary conditions can be expressed as
\begin{equation}
\Phi (\tau, \rho, z, \bar{z} ) = \sum_{l,m} \int d \omega d k \;  f_{\omega, k, l, m} (\tau, \rho, z, \bar{z}) \tilde{\Phi}(\omega,k,l,m)=\int d\omega dk f(\tau,\rho,z,\bar{z};k,\omega), \label{general}
\end{equation}
where $\tilde{\Phi}$ can be treated as coefficients and one can deduce them by applying the orthogonality relation of the basis 
\begin{equation}
    \int d\tau d\rho dz d\bar{z}\; w(\tau,\rho,z,\bar{z})\; f_{\omega,k,l,m}(\tau,\rho,z,\bar{z}) f_{\omega',k',l',m'}(\tau,\rho,z,\bar{z})=\delta_{ll'}\delta_{mm'}\delta(\omega-\omega')\delta(k-k')
    \end{equation}
with proper weight function $w$ deduced form the equation of motion. Sometimes it is convenient to do the sum over discrete variables and then absorb the coefficient term into the mode function therefore define the $(\omega,k)$ mode $f(\tau,\rho,z,\bar{z};k,\omega)$. In relation \eqref{general} we express the integrals abstractly; we will discuss how the domain of $(\omega,k)$ relates to boundary and regularity conditions below. 

We can also define a basis on spatial slices 
\begin{equation}
F_{k, l, m} (\rho, z, \bar{z}) = \phi_{l}(\rho, k)  Y^l_m(z,\bar{z})\label{tran}
\end{equation}
Any scalar satisfying the equation of motion can be expressed as
\begin{equation}
\Phi (\tau, \rho, z, \bar{z} ) = \sum_{l,m} \int d k \;  F_{k, l, m} (\rho, z, \bar{z}) \bar{\Phi}(\tau, k,l,m),\label{spatialde}
\end{equation}
where we have imposed the on-shell condition $\omega=k$ and reorganize the product of $\tilde{\Phi}(k,k,l,m)\psi(\tau,k)$ into $\bar{\Phi}(\tau,k,l,m)$. 

Analogously we can transform only in the time direction i.e. 
\begin{equation}
\Phi (\tau, \rho, z, \bar{z} ) =  \int d \omega  \;   \psi(\tau,\omega) \hat{\Phi} (\omega, \rho, z, \bar{z}), \label{tmode}
\end{equation}
where again we rewrite the data and make  $\sum_{lm}\tilde{\Phi}(\omega,\omega,l,m)F_{k,l,m}(\rho,z,\bar{z})$ into $\hat{\Phi}(\omega,\rho,z,\bar{z})$ . We will see later that those two are the most natural ways to read off the holographic data. 

\subsection{Explicit Modes}
\qquad In this section, we turn to the explicit solution of the differential equations above. These have been discussed in the literature \cite{deBoer:2003vf,Marolf:2006bk,cheung20174d,Liu:2021tif,raclariu2021lectures}, but here we will consider in further detail the role of regularity and boundary conditions. Together with stability, we will see that those requirements will impose constraints on the parameter $\omega$ or $k$.
\subsubsection*{Massless Fields}
Let us consider the differential equation in time. It is useful to consider first the case of a massless field, so that the equation reduces to
\begin{equation}
    \left(-3\frac{\partial_\tau}{\tau}-\partial^2_\tau+\frac{\omega^2}{\tau^2} \right) \psi(\tau, \omega)= 0\label{mode}
\end{equation}
after setting $M=0$ in \eqref{effective}. The generic solution takes the form
\begin{equation}
\psi(\tau, \omega) = \psi (\alpha_+) \tau^{-1 + \alpha_+} + \psi(\alpha_-) \tau^{-1 + \alpha_-}
\end{equation}
where $\alpha_{\pm}$ are the two roots of 
\begin{equation}
\alpha^2 = 1 + \omega^2 .\label{alpsq}
\end{equation}
Solutions are bounded $|\psi|<\infty$ at the null infinity $\tau\rightarrow \infty$ if either ${\rm Re}(\alpha_+) \leq 1$ or  ${\rm Re}(\alpha_-) \leq 1$. More precisely, states that are localised in the center and vanish at the boundary take the value $|{\rm Re}(\alpha)|<1$, called bound states. States which could propagate to the infinity and have non-zero contribution at null boundary are called scattering states. We will study them in a separate way.
\bigskip

{\it Scattering States}

\bigskip
For scattering states, we should have either ${\rm Re }(\alpha_+)=1$ or ${\rm Re }(\alpha_-)=1$. Thus all solutions that are finite at the infinity have the form
\begin{equation}
\alpha = 1 + i p 
\end{equation}
with $p$ real. We can write a general scattering state as 
\begin{equation}
\psi(\tau, p ) = \psi (p) e^{i p \ln \tau} 
\end{equation}
where $p$ is real, $\alpha^2 = (1-p^2) + 2 i p $ and $\omega^2 = 2 i p  - p^2$.  Clearly each such mode is not real. If $\alpha_+ = 1+ i p$, then the corresponding second root of $\eqref{alpsq}$ is $\alpha_ - = - (1 + i p)$; the latter mode is bounded as $\tau \rightarrow \infty$ but is not bounded as $\tau \rightarrow 0$. Thus for a given real value of $p$ the general solution takes the form 
\begin{equation}
\psi(\tau, p) = \psi_+ (p) \tau^{i p} + \psi_- (p)  \tau^{-i p - 2} \equiv \psi_{+} (p) f_+ (\tau,p) + \psi_- (p) f_- (\tau,p)\label{co1}
\end{equation}
To understand the orthogonality relation it is useful to first recall the standard relations for exponentials i.e. 
\begin{equation}
\int^{\infty}_{- \infty} d (\ln \tau) e^{i (p - q) \ln \tau}  =  \int^{\infty}_{0}  \frac{d \tau}{\tau} e^{i (p - q) \ln \tau}  = 2 \pi \delta (p - q)
\end{equation}
The latter is equivalent to 
\begin{equation}
\int^{\infty}_{0} d \tau w (\tau) f_{+} (\tau,p) f_- (\tau,q) = \delta ( p - q)
\end{equation}
where the weight function $w (\tau)= \tau$  is derived by expressing \eqref{mode} in standard Sturm-Liouville form i.e. 
\begin{equation}
\partial_{\tau} \left ( P(\tau) \partial_{\tau} \psi \right ) + Q(\tau) \psi = - \lambda w (\tau) \psi \label{sl}
\end{equation}
where $\lambda$ is the eigenvalue i.e. $\omega^2$ and the coefficient functions $(P(\tau),Q(\tau))$ follow from \eqref{mode}. 
\bigskip

{\it Bound States on Principal Series}

\bigskip

For bound states $|\psi|\rightarrow 0$ when $\tau\rightarrow \infty$, as we have mentioned, $\alpha$ should satisfy ${\rm Re}(\alpha_\pm)<1$. Here we are just interested in the special case such that $\alpha_\pm$ are chosen to be
\begin{equation}
    \alpha_\pm =\pm \;ip
\end{equation}
for $p \in \mathbb{R}$, then $\tau$ modes  $f_\pm$  will become
\begin{equation}
    f_+(\tau,p)=\tau^{-1+\alpha^+}=\frac{e^{ip\ln{\tau}}}{\tau},\qquad f_-(\tau,p)=\tau^{-1+\alpha^-}=\frac{e^{-ip\ln{\tau}}}{\tau}.
\end{equation}
Now we impose further restriction on $p$ so that make $p\geq 0$. This could always be done since $f_+(\tau,p)=f_-(\tau,-p)$ and one can treat such restriction as the reduction of the redundancy of the basis or the decomposition of the mode into positive and negative frequency components. For a generic function $\psi(\tau,p)$, we have the decomposition
\begin{equation}
    \psi(\tau,p)=\psi(p)f_+(\tau,p)+\psi^*(p)f_-(\tau,p),\label{co2}
\end{equation}
in which $\psi(p)$ are complex coefficients and $\psi(\tau,p)$ is now real. Given the weight function $w(\tau)=\tau$, one can check that
\begin{equation}
    \int_0^\infty d\tau w(\tau) f_+(\tau,p)f_-(\tau,q)=2\pi\;\delta(p-q)\label{orth1}
\end{equation}
and the relation
\begin{equation}
    \int_{0}^\infty d\tau w(\tau) f_+(\tau,p)f_+(\tau,q)=\int_{0}^\infty d\tau w(\tau) f_-(\tau,p)f_-(\tau,q)=2\pi\;\delta(p+q)=0.\label{orth2}
\end{equation}
Later, we will see that those states are dual to operators on the celestial sphere with scale dimension $\Delta$ satisfying
\begin{equation}
    \Delta=1+\alpha_+=1+ip,
\end{equation}
which is half of the principal series that forms the unitary representation of $SO(1,3)$ \cite{Dobrev:1977qv}. It is also worthwhile to note that the mode expansion \eqref{tmode} will become inverse Mellin transform if the $\tau$ modes take the form in \eqref{co2}.

Here we should note that the concept of bound and scattering are not absolute in a given physical theory. For example, one can also classify all the physical modes by the flux at the null boundary which behaves like $\tau^2 \psi^2$. If the scattering modes are defined as the modes that have non-zero flux at the boundary, the previous bound principal states will then become scattering states according to the new definition . The point is that, like quantum mechanics, we would like to emphasis that the behaviour of the field configuration at null infinity are related to $\alpha$, which could have explicit physical meaning based on the question we are interested in.
\subsubsection*{Massive Fields}
\qquad For non-zero mass the generic solution takes the form 
\begin{equation}
\psi(\tau, \omega) = \psi (\alpha_+) \frac{J_{\alpha_+} (M \tau)}{\tau}  + \psi(\alpha_-) \frac{J_{\alpha_-} (M \tau)}{\tau}
\end{equation}
where $\alpha_{\pm}$ are again the roots of \eqref{alpsq} and the first kind Bessel function is denoted as $J_\alpha$. Here we assume that $\alpha_{\pm}$ are generic complex numbers, in which case the two Bessel functions expressed in this form are manifestly linearly independent. For integer $\alpha$ the second solution will be expressed in the form of the second Bessel function $Y_{\alpha}$. Solutions that are bounded as $\tau \rightarrow 0$ have ${\rm Re} (\alpha ) \ge 1$ since $J_{\alpha} (M \tau) \sim \tau^{\alpha}$ as $\tau \rightarrow 0$. Using this limit of $J_{\alpha}(x)$ as $x \rightarrow 0$, the mode functions clearly reduce to those above as $M \rightarrow 0$. At large $\tau$, the Bessel function will be regular $J_\alpha\sim 1/\sqrt{\tau}$ at the null boundary thus it fits out intuition that the trajectory of massive particle will start at $i^-$ and end at $i^+$, which is the main difference from the massless case.  However the Bessel function will have the same orthogonality relations as \eqref{orth1} and \eqref{orth2} provided one has done the proper analytic continuation from the orthogonality relation for real $\alpha$. In the following sections, we will mainly use the massless solution as the example to perform the calculation while one should note that the results could be generalised to the massive case without conceptual obstacle. 

 As we have seen, the value of $\alpha(k)$ or equivalently $\Delta$ are often related to the behaviour of the solution near the light cone or null boundary. For example in section 3 of \cite{deBoer:2003vf}, they argued that the onshell action should be regular around the light cone thus, for the modes behaves as $\tau^{-1+\alpha}$, one requires
 \begin{equation}
   {\rm Re}(\alpha)\geq 0  
 \end{equation}
which is a weaker restriction than the boundedness condition ${\rm Re}(\alpha)\geq 1$. 

In section 2 of \cite{Marolf:2006bk}, the regularity of the solution is studied from the normalization point of view. The behaviour of the field at null boundary and light cone are both studied and it is argued that the solution should be oscillatory in order to make the mode normalizable. In our context, the oscillatory condition means that $\alpha$ should be complex, i.e. ${\rm Im}(\alpha)\neq 0$. Furthermore, Marolf also argued that the oscillatory fields should be separated into two parts. One is dynamical and it is normalizable according to the Klein-Gordon norm while the other part is not normalizable and is used to specify the boundary condition of the system. In additional to the Klein-Gordon norm, the other kind of paring between the oscillatory modes is also introduced in order to study the inner product structure between all the modes. 

More rigorous study of the asymptotic behaviour of the solution for Klein-Gordon equation in math literature are shown in \cite{vasy2013microlocal,baskin2015asymptotics,baskin2018asymptotics}. The boundedness of the solution for Schwarzschild case is shown in the gravity literature, so called Kay–Wald boundedness theorem \cite{kay1987linear} and one can see the review in \cite{Dafermos:2008en}. For Minkowski case, stability for the Einstein equation is first shown in \cite{christodoulou1993global}. Then for scalar-Einstein case when the matter field propagating on the asymptotically Minkowski background, the stability is also proofed provided the decay of the fields is under well controlled at the boundary \cite{lindblad2010global}, which leads to the constraints on the real part of the scale dimension i.e. ${\rm Re}(\alpha)$. This is similar to the study of Breitenlohner-Freedman bound for AdS spacetime \cite{Breitenlohner:1982bm,Breitenlohner:1982jf}. For the stability of Minkowski, a sharp bound for $\alpha$ is not found yet while, according to the above discussion, we summarise the possible range as \footnote{We should note that the upper bound comes from the boundedness of the modes  $|\psi|\leq \infty$ at the null infinity while for the stability of Minkowski space the condition will usually be stronger than the boundedness. For example, in the work \cite{lindblad2010global}, the decay behaviour of the field is required to be $|\psi|<\tau^{-1}$ thus the real part of $\alpha$ could only be zero after taking the lower bound into consideration. Here we choose to present the wider range for $\alpha$ although it not clear to us whether the value $0<|{\rm Re}(\alpha)|\leq 1$ are physical and stable or not. }
\begin{equation}
    0\leq{\rm Re}(\alpha)\leq 1,\qquad {\rm Im}(\alpha) \neq 0.\label{bound}
\end{equation}
However, given the fact that $\alpha$ has two solutions satisfying $\alpha_++\alpha_-=0$, we will have ${\rm Re}(\alpha_\pm)=0$ if one requires both of the modes $f_{\pm}(\tau,p)$ live in the bound \eqref{bound}. For other choices of $\alpha$, one of the two modes will be stable while the other will not. 

For the massive particles, we should note that the upper bound ${\rm Re}(\alpha)\leq 1$ will be relaxed since the requirement of regularity at the null boundary will not impose restrictions on the Bessel functions while it still not known whether the stability condition will introduce extra restrictions on the upper bound. The lower bound will not change since we have seen that the Bessel function will approach to the massless solution when $\tau\rightarrow 0$. 

\subsection{Radial Equation}\label{radialeq}

\qquad Now let us turn to the radial equation \eqref{effective}. It is important to distinguish between solutions to the equation for all radial values, and the asymptotic expansions from which the holographic dictionaries are constructed. The general solution to the radial equation can be written as
\begin{equation}
\phi_{l}(\rho; k) =  \phi(k) {\rm csch}\eta\; P_l^{\beta}({\rm coth}\eta) + \varphi(k) {\rm csch}\eta\; Q_l^{\beta}({\rm coth}\eta),
\end{equation}
in which $\rho=\sinh{\eta}$ and $(P,Q)$ are associated Legendre functions. Note that the range of $\eta$ is the same as that for $\rho$ i.e. $0 \le \eta < \infty$. The order of the function is given by 
\begin{equation}
\beta^2 = 1 + k^2 \label{beta}
\end{equation}
where here we do not assume that $\beta$ is real. In fact, since $\coth(\eta) \ge 1$ over the domain of interest, it is more useful to write the general solution in terms of the hypergeometry functions as shown in the appendix \ref{solutions} thus here it is convenient to choose the basis as 
\begin{equation}
\phi_{l}(\rho; k) =  \phi^+_{l} (k) {\rm csch}\eta\; P_l^{\beta_+}({\rm coth}\eta) +  \phi^-_{l} (k) {\rm csch}\eta\; P_l^{\beta_-}({\rm coth}\eta) ,
\end{equation}
where $\beta_{\pm}$ are the two (complex) roots of \eqref{beta}, with $(\beta_{+} + \beta_-) = 0$.  

To understand the regularity and orthogonality relations, it is useful to consider  first the $l=0$ solutions which can be written in terms of elementary functions as
\begin{equation}
\phi_0 (\rho;k) = \phi^+ (k) \frac{1}{\rho} ( \rho + \sqrt{\rho^2 + 1})^{\beta_+} + \phi^- (k) \frac{1}{\rho} ( \rho + \sqrt{\rho^2 + 1})^{\beta_-}
\end{equation} 
where $(\beta_{+} + \beta_-) = 0$. 
A mode is bounded as $\rho \rightarrow \infty$ provided that ${\rm Re} (\beta) \le 1$. However, no single mode is bounded as $\rho \rightarrow 0$. One can combine modes in a proper way to obtain fields $\phi_r(\rho;k)$ that are bounded as $\rho \rightarrow 0$:
\begin{equation}
\phi_r(\rho;k)=\frac{1}{\sqrt{\pi}\rho} \sinh \left (  \beta _+  \ln (\rho + \sqrt{\rho^2 + 1}) \right ).  
\end{equation}

The orthogonality condition for $l=0$ is obtained as above from writing the radial equation in Sturm-Liouville form \eqref{sl}, so that the coefficient and weight 
functions are given by
\begin{equation}
P(\rho) = \rho^2 (\rho^2 + 1)^{\frac{1}{2}} \qquad
w (\rho) = \frac{\rho^2}{ (\rho^2 + 1)^{\frac{1}{2}}}. 
\end{equation}
Therefore we have
\begin{equation}
\int_0^{\infty} d \rho w (\rho) {\cal F}^* (\rho;q) {\cal F} (\rho;p) = \delta(p-q) , \qquad p,\;q>0 \label{orth3}
\end{equation}
where $p\in \mathbb{R}$, $\beta=ip$ and 
\begin{equation}
{\cal F} ( \rho; p ) = \frac{1}{\sqrt{2\pi}\rho} ( \rho + \sqrt{\rho^2 + 1})^{ip}.
\end{equation}
Moreover, using the relation \eqref{orth3} one can also obtain the relation between the regular solutions written as
\begin{equation}
\int_0^{\infty} d \rho\; w (\rho) \;\phi_r^* (\rho; q) \phi_r (\rho,;p) = \delta(p-q).
\end{equation}
For the $l>0$ modes, the analysis of regularity is shown in the Appendix \ref{solutions} while the orthogonality relation will be hard to check and one can see the discussion in \cite{bielski2013orthogonality,laddha2022squinting} . 

Till now, we have discussed the modes with various choices of the value of $\alpha$ or $\beta$ and their corresponding physical interpretation while we should note that it is not clear which of them will form the necessary complete basis for the bulk fields and a generic principle to find out such a basis is still absent. Later we will see that different $k$-modes contribute to the correlation function living on the boundary celestial sphere in a different way according to the detail of the interaction. Here, we assume that given the detail of the theory a proper subset $\mathcal{P}$ of $k$ always exists that enable us to perform the mode decomposition thus the modes form a complete basis and the superposition principle will work. In the rest of this article, we will focus on the study of onshell fields therefore the condition $\alpha=\beta$ is automatically imposed. 
\section{Holography}\label{Dictionary}
\qquad The purpose of this section is to develop a detailed holographic dictionary between the bulk theory in asymptotically Minkowski spacetimes and the putative dual theory, associated with null infinity. We will develop the dictionary using the example of a test scalar field in the fixed Minkowski background. Our approach will be based on the principles of AdS/CFT \eqref{ads/cft}, i.e, writing a defining holographic relation of the form
\begin{equation}
    {\rm exp}\;\big(i S^{\rm ren}(\Phi)\big)=\Big\langle\; {\rm exp}\;\int_{\partial M}\mathcal{J}\;\mathcal{O}\;\Big\rangle_{QFT}.\label{flat/cft}
\end{equation}
Here $S(\Phi)$ is the action of the bulk theory with scalar field $\Phi$. Taking into account IR divergences, we will need to renormalise this action and $S^{\rm ren}(\Phi)$ is the renormalised version of $S(\Phi)$; an important part of this section will be establishing the principles underlying the renomalisation procedure. On the right hand side of \eqref{flat/cft} we denote $\mathcal{J}$ and $\mathcal{O}$ as the source and operator in the quantum field theory at the boundary. Again this should be viewed as a renormalised expression.

Following the story of the construction of the dictionary for AdS/CFT \cite{Witten:1998qj,Freedman:1998tz,Klebanov:1999tb}, here we are also going to specify the source and operator by decomposing the data of the bulk field $\Phi$ into coefficients when doing the expansion at the boundary. It turns out that we need two series of operators $\{\mathcal{J},\mathcal{O}\}$ and $\{\Tilde{\mathcal{J}},\tilde{\mathcal{O}}\}$ on the celestial sphere in order to reconstruct the bulk field by checking the renormalised action specifically and the way that they are coupled is determined by the causal and dynamical structure of the bulk theory. The new feature for the flat/CFT dictionary is that we are reducing two spacetime dimensions at once and the dictionary is built between the bulk theory with the  notion of time and the boundary Euclidean theory on the sphere thus the factor $i$ plays an important role here when considering the emergence of time and the unitarity of the CFT.
\subsection{Holographic Dictionary}

\qquad We begin by reviewing the usual holographic dictionary for scalar fields on Euclidean AdS$_3$. Using the same coordinates for Euclidean AdS$_3$ as shown in \eqref{milen} i.e. 
\begin{equation}
ds_{\rm AdS_3}^2 = g_{ij} dx^i dx^j=\left(\frac{d\rho^2}{1+\rho^2}+2\rho^2 \gamma_{z\bar{z}}dzd\bar{z} \right), \label{ads3}
\end{equation} 
the boundary is at $\rho \rightarrow \infty$ and the boundary metric is manifestly spherical. Now consider a massive scalar field with action
\begin{equation}
S_{\rm AdS_3} = \frac{1}{2} \int d^3 x \sqrt{g} \left ( (\partial \varphi)^2 + m^2 \varphi^2 \right ),
\end{equation}
where $g$ is the determinant of the Euclidean AdS$_3$ metric above. The onshell action is thus
\begin{equation}
S^{\rm onshell}_{\rm AdS_3} = \frac{1}{2} \int_{\partial AdS_3} d \Sigma^i \varphi \partial_i \varphi 
\end{equation}
where $d\Sigma^i=dS\frac{1}{\sqrt{g}}\partial_j \sqrt{g} g^{ij}$ and $dS$ is the volume form on the surface at the cut off $\rho=R$. In terms of the coordinate \eqref{ads3}, it takes the form $d\Sigma^\rho=d^2z \;\gamma_{z\bar{z}}R^2\sqrt{1+R^2}$ while the other two components will vanish $d\Sigma^z=d\Sigma^{\bar{z}}=0$ since the sphere is compact. The asymptotic expansion of an onshell field takes the form
\begin{equation}
\varphi (\rho,z) = \rho^{ \Delta - 2} \left (\varphi (z) + \cdots \right) + \rho^{- \Delta} \left ( \tilde{\varphi} (z) + \cdots \right )
\end{equation}
where $\varphi(z)$ is the source for the dual operator ${\cal O}_{\varphi} (z)$ of dimension $\Delta$, where $m^2 = \Delta (\Delta - 2)$. When $\Delta$ is integral the asymptotic expansions contain logarithmic terms, which are related to the contact terms in two point functions discussed below. 

One uses the asymptotic expansion of the onshell field to compute the explicit value of the regulated onshell action, from which one can construct covariant counterterms and the renormalised action
\begin{equation}
S^{\rm ren}_{\rm AdS_3} = {\cal L}_{\rho \rightarrow \infty} \left ( S^{{\rm onshell}}_{{\rm AdS_3}} + S^{\rm ct}_{\rm AdS_3} \right )
\end{equation} 
The covariant counterterms are of the form
\begin{equation}
S^{\rm ct}_{\rm AdS_3} = - \frac{1}{2} (\Delta - 2 ) \int_{\partial AdS_3} d^2 x \sqrt{h} \varphi^2 + \cdots
\end{equation}
where $h$ is the determinant of the induced metric at the boundary. In the chosen coordinates \eqref{ads3}, it takes the form $h=R^2 \gamma_{z\bar{z}}=R^2\Omega_2(z)$. 

In terms of the complex AdS coordinate \eqref{ads3}, the ${\rm AdS}_3/{\rm CFT}_2$ dictionary the can be written as
\begin{equation}
    {\rm exp}\;\Big(- S_{{\rm AdS}_{3}}(\Phi)\Big)=\Big\langle\; {\rm exp}\;-\int_{S^2} d^2z \Omega_2(z) \mathcal{J}(z)\;\mathcal{O}(z)\;\Big\rangle,\label{ads/cftc}
\end{equation}
where $\mathcal{J}(z)\sim \Omega_2^{\frac{\Delta-2}{2}}(z)\varphi(z)$ \footnote{It is easier to keep track of the weight if one chooses to use the expansion $\phi(\rho,z)=(\rho \Omega^{\frac{1}{2}}_2(z))^{\Delta-2}\varphi(z)+(\rho \Omega^{\frac{1}{2}}_2(z))^{-\Delta}\tilde{\varphi}(z)$ and one should note the relation $\rho (\frac{\Omega(z)}{2})^{\frac{1}{2}}= \frac{1}{t}$ between the complex and Poincar$\rm \acute{e}$ coordinates.} is the corresponding source and the expectation value of the dual operator is then defined as the variation of the renormalised action with respect to the source
\begin{equation}
    \langle {\cal O}_\varphi (z) \rangle _{S^2}=\frac{1}{\Omega_2(z)}\frac{\delta S^{\rm ren}(\Phi)}{\delta \mathcal{J}(z)},
\end{equation}
in which $\langle \cdots \rangle_{S^2}$ means that the operator is inserted on the celestial sphere. From the above definition, one can deduce that the one-point function is then proportional to the function $\Omega_2^{-\frac{\Delta}{2}}(z)\tilde{\varphi}(z)$, i.e. $\langle\mathcal{O}_\varphi \rangle_{S^2}\sim \Omega_2^{-\frac{\Delta}{2}}(z)\Tilde{\varphi}(z)$. In this article, we are interested in the operator on the two dimensional plane denoted as $M_2$ thus the correlations functions are related by the conformal weight $\langle \mathcal{O}_\varphi(z)\rangle\sim \Omega_2^{\frac{\Delta}{2}}(z)\langle \mathcal{O}_{\varphi}(z)\rangle_{S^2}$. Therefore, after taking the renormalisation factor into consideration, one has the relation
\begin{equation}
\langle {\cal O}_\varphi (z) \rangle = 2 (1 - \Delta) \tilde{\varphi} (z) + C(\varphi)
\end{equation} 
for the operator on the complex plane and the source now becomes $\mathcal{J}(z)=\varphi(z)$. Here the function $C(\varphi)$ denotes contributions to the one point correlation function that are expressed in terms of the source; such contributions arise whenever $\Delta$ is integral and its exact form depends on the regularization scheme. As usual the two point function can be obtained by functionally differentiating with respect to the source $\varphi(z)$ i.e. 
\begin{equation}
\langle {\cal O}_\varphi (z) {\cal O}_\varphi (z') \rangle = - 2 (1 - \Delta)  \frac{\delta \tilde{\varphi} (z)}{\delta \varphi(z')} + \cdots
\end{equation} 
where the ellipses contribute only to contact terms in the correlation function and the renormalisation factor $2(\Delta-1)$ can be deduced by the study of bulk-boundary propagator which is briefly reviewed in the Appendix \ref{coordinates}.

Given the bulk-boundary propagator $K(\rho,z;z')$, a generic regular field in the bulk with boundary behaviour $\varphi(\rho,z)\sim \rho^{\Delta-2}\varphi(z)$ can be expressed as 
\begin{equation}
    \varphi(\rho,z)=\int_{M_2}  dz'd\bar{z}' \;K(\rho,z;z')\varphi(z'), \label{fieldsource}
\end{equation}
in which we have transformed the boundary to the plane and the source becomes $\mathcal{J}(z)=\varphi(z)$. With the help of the AdS/CFT propagator, one can deduce the CFT two-point function in a quick way. For example, the ${\rm AdS_3}$ onshell action can be written as
\begin{eqnarray}
    S^{\rm onshell}_{\rm AdS_3}&=&\frac{1}{2}\int_{M_2}d^2z\;R^2\sqrt{1+R^2}\;(\varphi(\rho,z)\partial_\rho \varphi(\rho,z))_{\rho=R}\\
    &=&-\frac{\Delta}{2\pi}\int_{M_2}\int_{M_2} d^2z d^2z'  \;\frac{\varphi(z)\varphi(z')}{|z-z'|^{2\Delta}},
\end{eqnarray}
in which in the second line we have used the expression \eqref{fieldsource} and the contraction relation \eqref{contract} for the propagators. Following the similar procedure, one can also deduce $S^{\rm ct}_{\rm AdS_3}$ and we have
\begin{eqnarray}
    S^{\rm ct}_{\rm AdS_3}&=&-\frac{1}{2}(\Delta-2)\int_{M_2}d^2z\;R^2\;(\varphi(\rho,z) \varphi(\rho,z))_{\rho=R}\\
    &=&-\frac{\Delta-2}{2\pi}\int_{M_2}\int_{M_2} d^2z d^2z'  \;\frac{\varphi(z)\varphi(z')}{|z-z'|^{2\Delta}},
\end{eqnarray}
therefore, according to the AdS/CFT dictionary, the renormalised two-point function now becomes 
\begin{equation}
    \langle \mathcal{O}(z)\mathcal{O}(z')\rangle=-\frac{\delta^2 S^{\rm ren}_{\rm AdS_3}}{\delta\varphi(z)\delta\varphi(z')}=\frac{c_\Delta}{|z-z'|^{2\Delta}},\label{2-pt}
\end{equation}
where $c_\Delta$ takes the value \footnote{For the massive case, we have $c_\Delta=\frac{2(\Delta-1)N_{\Delta}}{\pi}$. We should also note that the contact terms in the two-point function coming from the approximation \eqref{bulk-boundary} and \eqref{contract} have been ignored here. }
\begin{equation}
    c_\Delta=\frac{2(\Delta-1)}{\pi}.
\end{equation}
\subsection{Holography Dictionary for Milne}
\qquad In this section we turn to scalar fields in the Milne coordinates then proceed to perform the holography renormalisation for Minkowski spacetime. The action for the massive scalar field is 
\begin{equation}
S = \frac{1}{2} \int_{0}^{\infty} d \tau \int_0^\infty d\rho \int d z d \bar{z} \sqrt{-G} \left ( (\partial\Phi)^2 + M^2 \Phi^2 \right ),
\end{equation}
in which $G$ is given by \eqref{milen} together with scalar fields $\Phi$ and we have restricted the integration region to $\mathcal{A}^+$. As usual we can express the onshell action as the exact term 
\begin{equation}
S^{\rm onshell} = \frac{1}{2} \int_0^\infty d \tau \int_0^\infty d\rho \int d z d \bar{z} \sqrt{-G} D^{\mu} ( \Phi \partial_{\mu} \Phi ),\label{boundary1}
\end{equation}
which can be expressed as boundary terms thus we have $D^\mu=\frac{1}{\sqrt{-G}}\partial_\nu \sqrt{-G} G^{\nu\mu}$. The philosophy of the celestial holography approach is to foliate the spacetime with spacelike surfaces, and throughout this section we will work in this approach, analysing divergences at the spatial boundaries of each slice. 

Accordingly, let us focus on the radial boundary as $\rho \rightarrow \infty$. Using the Milne form of the metric the onshell boundary terms are 
\begin{equation}
S^{\rm onshell} = \frac{1}{2} \int_0^\infty d \tau \tau \int_{\partial AdS_3} d\Sigma^i \Phi (\tau, x^i) \partial_i \Phi (\tau, x^i)\label{boundary2}
\end{equation} 
where the second integral is expressed in terms of the boundary of the Euclidean AdS$_3$ metric \eqref{ads3}. Here we should note that, strictly speaking, the value of onshell action shown in \eqref{boundary1} and \eqref{boundary2} are not the same since we have ignored the integral over spatial direction at the fixed hyperboloids $\tau=0$ and $\tau=+\infty$. After taking the other Milne wedge $\mathcal{A}^-$ into consideration, the difference is then determined by the integral over $\Phi(\tau=\pm\infty,\rho,z,\bar{z})\partial_\tau\Phi(\tau=\pm\infty,\rho,z,\bar{z})$ in which $\Phi(\tau=\pm\infty,\rho,z,\bar{z})$ are the initial and final data imposed for a given physical system since $\tau=\pm \infty$ are null boundaries of Minkowski space. If one proposes that the initial and final states for the physical system are vacuum, then we have $\Phi(\tau=\pm\infty,\rho,z,\bar{z})=0$ thus there will be no difference between \eqref{boundary1} and \eqref{boundary2}. For scattering processes, the initial and final states are the in and out going states while one can assume the difference will contribute to the action in a small and finite way therefore leads to a proper $i\epsilon$ perscription of the quantum theory \cite{weinberg1995quantum}. One can see the formal treatment of the integral along the null boundaries and the renormalisation of the phase space in the work \cite{Compere:2011ve,Costa:2012fm,Capone:2023roc}. Here we will only study the onshell action in the form of \eqref{boundary2} and the explicit expression for this is
\begin{equation}
S^{\rm onshell} = \frac{1}{2} \int_0^\infty d \tau \tau \int_{\partial AdS_3} d \Omega_2 R^2  (1 + R^2)^{\frac{1}{2}} \left ( \Phi (\tau, \rho, z,\bar{z})  \partial_\rho \Phi (\tau, \rho, z, \bar{z}) \right )_{\rho = R}
\end{equation} 
where the boundary is regulated at $\rho =R$ and $d \Omega_2$ is the integration measure over the unit two sphere. 

Given the onshell solution, following the expansion \eqref{general}, we can further decompose it into the $k$ mode components by introducing the $k$ mode function $f(\tau,\rho,z,\bar{z};k)\delta(\omega-k)=f(\tau,\rho,z,\bar{z};k,\omega)$ thus we have
\begin{equation}
    \Phi(\tau,\rho,z,\bar{z})=\int_{ \mathcal{P}} dk f(\tau,\rho,z,\bar{z};k)\label{kmode}
\end{equation}
and then we can use such decomposition of fields to transform the onshell action into $k$  mode space after rewriting all the fields in the action in terms of $f$. More precisely, we can define the $(k,k')$ mode of the action
\begin{equation}
    S^{\rm onshell}(k,k'):=\frac{1}{2}\int_0^\infty\tau d\tau \int_{\partial AdS_3} d\Omega_2 \;R^3 (f(\tau,\rho,z,\bar{z};k)\partial_\rho f(\tau,\rho,z,\bar{z};k'))_{\rho=R}
\end{equation}
and one can check at large $R$ we have
\begin{equation}
    S^{{\rm onshell}}=\int_{\mathcal{P}} dk\int_{\mathcal{P}'} dk'\; S^{\rm onshell}(k,k'),
\end{equation}
where the double integral over the set $\mathcal{P}$ come from the fact that the onshell action for free particles are quadratic in terms of $\Phi$. Moreover, we can treat the $(k,k')$ mode of the action $S^{\rm onshell}(k,k')$ as the onshell action which describes the interaction between a pair of modes $(k, k')$. Later we will see that $S^{\rm onshell}(k,k')$ is proportional to the delta function if the domain of the integral over $k$ takes the value such that $\beta_+(k)=i\mathbb{R}^+$ thus we have
\begin{equation}
   S^{\rm onshell}(k,k') =\delta(k-k') S^{\rm onshell}  (k,k)
\end{equation}
and for simplicity we denote $S^{\rm onshell}(k,k)$ as $S^{\rm onshell}(k)$. In such convention, the onshell action then can be expressed as 
    \begin{equation}
   S^{\rm onshell}=\int_{\mathcal{P}} dk \;S^{\rm onshell}  (k),
\end{equation}
which will be used as the standard form of the $k$ mode decomposition of the action for free particles.

Before performing the renormalisation on $S^{\rm onshell}(k)$, let us consider asymptotic solutions of the equation \eqref{effective} as $\rho \rightarrow \infty$. The generic form for the asymptotic solution 
\begin{eqnarray}
\phi_{l}(\rho ; k) &=& \phi_l (\rho ; \beta_+(k)) + \phi_l(\rho; \beta_-(k)) \\
& \equiv &   \rho^{\beta_+ - 1}  \left ( \phi^+_{l} (k) + {\cal O}\left ( \frac{1}{\rho^2} \right ) \right ) 
+ \rho^{\beta_- - 1} \left ( \phi^-_{l} (k) + {\cal O}\left ( \frac{1}{\rho^2} \right ) \right )  \nonumber 
\end{eqnarray}
where $(\beta_+ + \beta_- ) = 0$ and without loss of generality we will assume that ${\rm Re}( \beta_+) \ge {\rm Re} (\beta_-)$. Instead of using $l$ modes on the sphere 
we can express a general solution for the spatial part of the scalar for fixed $k$ as
\begin{equation}
\phi(\rho, z, \bar{z}; k) = \phi(\rho, z, \bar{z}; \beta_+) + \phi(\rho, z, \bar{z}; \beta_-) \label{spat1}
\end{equation} 
where the asymptotics of each solution are of the form
\begin{equation}
\phi(\rho, z, \bar{z}; \beta_\pm) = \rho^{\beta_\pm - 1}  \left ( \phi^\pm (z,\bar{z}; k) + {\cal O}\left ( \frac{1}{\rho^2} \right ) \right ). \label{spat2}
\end{equation}
Combining modes of a fixed value of $k$ we obtain 
\begin{eqnarray}
f(\tau,\rho,z,\bar{z}; k) &=& f_+(\tau,k)\phi(\rho,z,\Bar{z};k)+f_-(\tau,k)\Tilde{\phi}(\rho,z,\bar{z};k)\\
&=&  \tau^{\beta_+ - 1} \phi (\rho, z, \bar{z}; \beta_+) +  \tau^{\beta_{-} -1 } \tilde{\phi }(\rho, z, \bar{z}; \beta_+) \label{spit1} \\
&& +  \tau^{\beta_+ - 1} \phi (\rho, z, \bar{z}; \beta_-) + \tau^{\beta_{-} -1 }  \tilde{\phi} (\rho, z, \bar{z}; \beta_-) \nonumber 
\end{eqnarray}
where the fields $\tilde{\phi} (\rho, z, \bar{z}; \beta_\pm)$ have the properties \eqref{spat1} and \eqref{spat2} and well see the explicit expression for them in the next section.

Now let us return to the four-dimensional $k$ mode action. The regulated action for modes of fixed $k$ contains the terms
\begin{eqnarray}
S^{\rm onshell}  (k) &=& \frac{1}{2} \int_0^\infty d \tau \tau \int_{\partial AdS_3} d \Omega_2 \left (  (\beta_+ - 1)R^{2 \beta_+} \Phi_s (\tau, z, \bar{z}; k)^2 + ( \beta_- - 1)R^{2 \beta_-} \Phi_v (\tau, z, \bar{z}; k)^2  \right .\nonumber  \\
&& \qquad \qquad \qquad \qquad  \left . - 2 \Phi_s (\tau, z, \bar{z}; k) \Phi_v (\tau, z, \bar{z}; k) + \cdots \right ), \label{on4}
\end{eqnarray}
where the boundary of the AdS slice is regulated as $\rho = R$ and the ellipses denote terms that are suppressed by at least $1/R^2$. We introduce a shorthand notation for the combinations of terms in the asymptotic radial expansions:
\begin{eqnarray}
\Phi_s (\tau, z, \bar{z}; k) &=& \tau^{\beta_+ - 1 } \phi^+ (z,\bar{z}; k) + \tau^{\beta_- -1}  \tilde{\phi}^+ (z,\bar{z}; k) \\
\Phi_v (\tau, z, \bar{z}; k) &=& \tau^{\beta_+ - 1 } \phi^- (z,\bar{z}; k) + \tau^{\beta_- -1}  \tilde{\phi}^- (z,\bar{z}; k) \nonumber
\end{eqnarray}
Let us suppose that ${\rm Re} (\beta_+) > 0$, in which case ${\rm Re} (\beta_-) < 0$. In this case the first term in \eqref{on4} will be divergent as $R \rightarrow \infty$, but the second term will vanish; all power law divergences will be of the form $R^{2 \beta_+ - 2n}$ with $n$ an integer. 

As above, we can remove divergences with counterterms. These counterterms should be expressed in terms of quantities that are intrinsic to the regulated boundary, and they should be covariant with respect to the bulk diffeomorphism at $\rho=0$ thus make $f(\tau,R,z,\bar{z})$ transform as a scalar field. Here in fact, the background metric already uses a preferred slicing of the four-dimensional metric, i.e. a specific coordinate choice for time, and therefore we would not expect the counterterms to preserve full three-dimensional covariance of the boundary. In practice this means that the counterterms are expressed in the form
\begin{eqnarray}
S^{\rm ct}(k) &=&\int_0^\infty d \tau \frac{1}{\tau}\int_{\partial AdS_3}d^2z \sqrt{-\bar{\gamma}} \left (a_1(k) f (\tau,R,z,\bar{z})^2 + a_2(k) (\partial_z \partial_\tau f (\tau,R,z,\bar{z}))^2 + \cdots \right )\nonumber\\
&=& \int_0^\infty d \tau \tau \int_{\partial AdS_3} R^2 d\Omega_2 \left (a_1(k) f (\tau,R,z,\bar{z})^2 + a_2(k) (\partial_z\partial_\tau f (\tau,R,z,\bar{z}))^2 + \cdots \right )
\end{eqnarray}
where $\bar{\gamma}_{\tau\tau}=-1,\;\bar{\gamma}_{z\bar{z}}=R^2\tau^2 \gamma_{z\bar{z}}$ is the induced metric on the boundary of Milne wedge at $\rho=R$ (with the curvature radius being independent of $\tau$) and the derivative $\partial_z$ only acts on the celestial sphere. As we can see, the covariant of the bulk diffeomorphism at the surface $\rho=R$ shown in the first line is broken by fixing the gauge of coordinates in the second line.  By construction these counterterms will remove the divergences because the analytic structure on the celestial sphere is precisely as described above for AdS$_3$/CFT$_2$. Indeed, matching with the dictionary above one obtains
\begin{equation}
\Delta_k = 1 + \beta_+  \label{scale}
\end{equation}
and furthermore one can deduce the factor corresponding to the first term in the counterterm to be 
\begin{equation}
    a_1(k)=\frac{2-\Delta_k}{2}
\end{equation}
thus the finite terms in the renormalized action include
\begin{eqnarray}
&&S^{\rm ren} (k) = -\beta_{+} \int_0^\infty d \tau \tau \int_{\partial AdS_3} d \Omega_2 \;R^2\; \Phi_s (\tau, z, \bar{z}; k) \Phi_v (\tau, z, \bar{z}; k) \\
&=& -\beta_+ \int_0^\infty d \tau \tau \int_{\partial AdS_3} d \Omega_2 (\tau^{\beta_+ - 1 } \phi^+ (z,\bar{z}; k) + \tau^{\beta_- -1}  \tilde{\phi}^+ (z,\bar{z}; k) )
(\tau^{\beta_+ - 1 } \phi^- (z,\bar{z}; k) + \tau^{\beta_- -1}  \tilde{\phi}^- (z,\bar{z}; k)). \nonumber
\end{eqnarray}
As we have discussed, there would be additional finite terms in the action if $\beta_+$ were to be real and integer valued, but this is not the case of interest here. According to the dictionary given in \eqref{flat/cft}, one can see that $\beta_+$ should be a pure imaginary number $\beta_+\in i\mathbb{R}$ in order to make sure that CFT correlation functions given by the right hand side of \eqref{flat/cft} are real. Moreover, given the relation \eqref{scale}, we know that the scale dimension of the operator on the celestial sphere should take the value on the principal series $\Delta_k=1+i\mathbb{R}$\footnote{In fact $\Delta_k=1+i\mathbb{R}^+$ if $\beta_+$ takes the value in $i\mathbb{R}^+$ and we assume that such $k$ modes will form the necessary complete basis when one performs the mode decomposition following the discussion in section \ref{Mode}. There are also shadow operators given by the shadow transformation $\Delta_k\rightarrow 2-\Delta_k$ so that the value of scale dimension will cover the whole principal series. From the bulk side, this corresponds to doing the Legendre transformation of the action so that the role of source and vev are switched.}. 

Using the orthogonality relations for the $\tau$ eigenfunctions \eqref{orth1}, \eqref{orth2}  we can explicitly compute the $\tau$ integrals as
\begin{equation}
S^{\rm ren} (k) =- \beta_+ \int_{\partial AdS_3} d \Omega_2 ( \phi^+ (z,\bar{z}; k) \tilde{\phi}^- (z,\bar{z}; k)  + \tilde{\phi}^+ (z,\bar{z}; k)  \phi^- (z,\bar{z}; k) )+\cdots
\end{equation}
and one can also see that the $\delta(k-k')$ will come out if one choose to use $S^{\rm onshell}(k,k')$ rather than $S^{\rm onshell}(k)$. From this expression we can read off that there are two operators of dimension $\Delta_k$ with corresponding expectation values and sources:
\begin{eqnarray}
\langle {\cal O} (z, \bar{z}; k) \rangle  &=& -2i\beta_+  \phi^- (z, \bar{z}; k) \qquad \mathcal{J}(z,\bar{z};k)=\tilde{\phi}^+(z,\bar{z};k)  \label{vev1} \\
\langle \tilde{\cal O} (z, \bar{z}; k) \rangle  &=& -2i\beta_+ \tilde{\phi}^- (z, \bar{z}; k) \qquad \tilde{\mathcal{J}}(z,\bar{z};k)={\phi}^+(z,\bar{z};k) \nonumber
\end{eqnarray}
These two operators have the same two dimensional CFT scaling dimension, but are associated with different evolution in the $\tau$ direction. 

A generic massless field $\Phi$ will be expressed as an integral over $k$, with the corresponding renormalized action being 
\begin{eqnarray}
S^{\rm ren} &=& \int_\mathcal{P} dk S^{\rm ren} (k) \\
&=&- \int_\mathcal{P} dk \; \beta_+ (k) \int_{\partial AdS_3} d \Omega_2 ( \phi^+ (z,\bar{z}; k) \tilde{\phi}^- (z,\bar{z}; k)  + \tilde{\phi}^+ (z,\bar{z}; k)  \phi^- (z,\bar{z}; k) ) \nonumber
\end{eqnarray}
The field $\Phi$ is thus dual to two continuous series of operators, labelled by $k$, whose sources and expectation values are given above in \eqref{vev1}. 
\subsection{Correlation Functions}\label{2point}

\qquad In this section we are going to study correlation functions in the context of flat/CFT in a more precise way.  Propagators for free fields $\langle\mathcal{O}\mathcal{O}\rangle$, $\langle\tilde{\mathcal{O}}\tilde{\mathcal{O}}\rangle$ are deduced and also represented in the language of diagrams. For higher point correlation functions in an interacting theory, the interactions are described by internal vertices of the diagrams. We use the $\Phi^3(X)$ interaction as an example to see how the operators of different scale dimensions are coupled with each other.
\begin{figure}[h!]
    \centering
    \begin{tikzpicture}
\draw(0,1) ellipse (1 and 0.5);
\draw(0,-1) ellipse (1 and 0.5);
\filldraw  (-1,1) circle (1.5pt);
\filldraw  (-1,-1) circle (1.5pt);
\draw[thick][->](-1,1)--(-1,0);
\draw[thick](-1,-1)--(-1,0);
\draw(-1.5,1) node{$\mathcal{O}$};
\draw(0,1) node{+};
\draw(0,-1) node{-};
\draw(-2,0) node{$\langle\mathcal{O}\mathcal{O}\rangle$};
\draw(2,0) node{$\langle\tilde{\mathcal{O}}\tilde{\mathcal{O}}\rangle$};
\draw(-1.5,-1) node{$\mathcal{J}$};
\filldraw  (1,1) circle (1.5pt);
\filldraw  (1,-1) circle (1.5pt);
\draw[thick](1,1)--(1,0);
\draw[thick][->](1,-1)--(1,0);
\draw(1.5,1) node{$\tilde{\mathcal{J}}$};
\draw(1.5,-1) node{$\tilde{\mathcal{O}}$};
\end{tikzpicture}    
    \caption{Propagators between two copies of operators are illustrated in the figure. Each disk represents one copy of ${\rm AdS }_3$ hyperboloid with $S^2$  boundary drawn as a circle. + and - represent that the operators are obtained from the decomposition of the $\tau^{\beta_+-1}$ or $\tau^{\beta_--1}$ modes.}
    \label{propagators}
\end{figure}
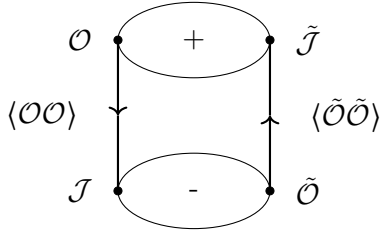

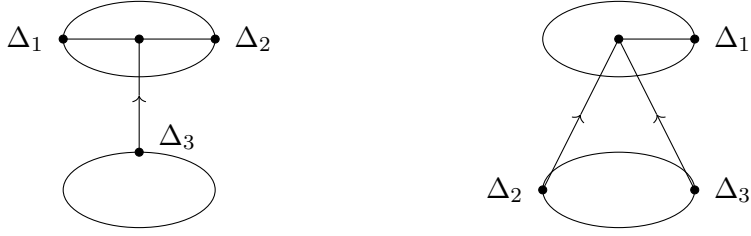
\begin{figure}
    \centering
    \begin{tikzpicture}
\draw(0,1) ellipse (1 and 0.5);
\draw(-1.5,1) node{$\Delta_1$};
\draw(1.5,1) node{$\Delta_2$};
\draw(0.5,-0.3) node{$\Delta_3$};
\draw(0,-1) ellipse (1 and 0.5);
\filldraw  (-1,1) circle (1.5pt);
\filldraw  (0,1) circle (1.5pt);
\filldraw  (0,-0.5) circle (1.5pt);
\draw(-1,1)--(0,1);
\draw(1,1)--(0,1);
\draw[->](0,-0.5)--(0,0.25);
\draw(0,0.25)--(0,1);
\filldraw  (1,1) circle (1.5pt);
\end{tikzpicture}
\hspace{6.0em}
\begin{tikzpicture}
\draw(0,1) ellipse (1 and 0.5);
\draw(0,-1) ellipse (1 and 0.5);
\draw(1.5,1) node{$\Delta_1$};
\filldraw  (-1,-1) circle (1.5pt);
\filldraw  (1,1) circle (1.5pt);
\filldraw  (0,1) circle (1.5pt);
\draw[->](-1,-1)--(-0.5,0);
\draw(-0.5,0)--(0,1);
\draw[->](1,-1)--(0.5,0);
\draw(0.5,0)--(0,1);
\draw(1,1)--(0,1);
\draw(-1.5,-1) node{$\Delta_2$};
\draw(1.5,-1) node{$\Delta_3$};
\filldraw  (1,-1) circle (1.5pt);
\end{tikzpicture}    
 \caption{For $\lambda_3\Phi^3(X)$ interaction, two kinds of $k$-mode contribution for the three point function $\langle\tilde{\mathcal{O}}\tilde{\mathcal{O}}\tilde{\mathcal{O}}\rangle$ are shown in the figure. The diagram on the left represents mode contribution like $\phi\phi\tilde{\phi}$ while the mode contribution like $\phi\tilde{\phi}\tilde{\phi}$ are shown on the right. }
\label{3pt-}
\end{figure}

Following the previous approach in the context of AdS/CFT, we again choose to decompose the bulk fields into the  bulk-boundary propagator
\begin{eqnarray}
     \Phi(\tau,\rho,z,\bar{z})&=&\frac{1}{2\sqrt{2}}\int_\mathcal{P} dk\int_{M_2} dz'd\bar{z}' \;\big(\tau^{\beta_+-1} K(\rho,z;\; z',\beta_+)\phi^+(z,\bar{z};k)\nonumber\\
     && \qquad\qquad\qquad+\tau^{\beta_--1} K(\rho,z;\; z',\beta_+)\tilde{\phi}^+(z,\bar{z};k)\big ),
\end{eqnarray}
in which $\partial AdS_3=M_2$ and $\phi^+(z,\bar{z};k)$, $\tilde{\phi}^+(z,\bar{z};k)$ can be treated as a pair of sources on the boundary as introduced in \eqref{vev1}. The extra factor $1/\sqrt{2}$ is introduced here to make the renormalisation factor the same as AdS/CFT case, since we have two modes from the decomposition, while one can also treat it as the rescaling of the propagator. Given such expression, from the onshell action 
\begin{eqnarray}
    S^{\rm onshell}(\Phi)= -\frac{1}{2}\int_{0}^{\infty}d\tau \int_{M_2} d^2z \;R^3\;(\Phi(\tau,\rho,z,\bar{z})\partial_\rho\Phi(\tau,\rho,z,\bar{z}))_{\rho=R}
\end{eqnarray}
we have the $k$-mode component 
\begin{equation}
    S^{\rm onshell}(k)=\frac{\Delta_k}{2\pi}\int_{M_2}\int _{M_2} d^2zd^2z'\frac{\phi^+(z,\bar{z};k)\tilde{\phi}^+(z',\bar{z}';k)}{|z-z'|^{2\Delta_k}},
\end{equation}
in which we have integrated out the $\tau$ variable and the orthorgonal relations for the $\tau$-modes are also applied. After performing the holographic renormalisation introduced in the previous section, the counterterm is then deduced to be 
\begin{eqnarray}
    S^{\rm ct}(k)=-\frac{1}{2}(\Delta_k-2)\int_{0}^{\infty}d\tau \int_{M_2} d^2z \;R^2\;(f(\tau,\rho,z,\bar{z})f(\tau,\rho,z,\bar{z}))_{\rho=R}+\cdots
\end{eqnarray}
with $k$-mode
\begin{equation}
    S^{\rm ct}(k)=-\frac{1}{2\pi}(\Delta_k-2)\int_{M_2}\int_{M_2} d^2zd^2z' \frac{\phi^+(z,\bar{z};k)\tilde{\phi}^+(z',\bar{z}';k)}{|z-z'|^{2\Delta_k}}.
\end{equation}
Given the flat/CFT dictionary, in order to obtain the two-point function of the operator $\mathcal{O}$, we need to do the variation with respect to the corresponding source $\mathcal{J}=\tilde{\phi}^+$ twice therefore get 
\begin{equation}
 \langle \mathcal{O}(z,\bar{z};k)\mathcal{O}(z',\bar{z}';k)\rangle= \frac{i\delta^2 S^{\rm ren}(k)}{\delta \tilde{\phi}^+(z)\tilde{\phi}^+(z')} =\int_{M_2} d^2z''\frac{\delta\phi^+(z'',\bar{z}'';k)}{\delta \tilde{\phi}^+(z,\bar{z};k)}\;\frac{c_k}{|z''-z'|^{2\Delta_k}},
\end{equation}
in which $c_k=2i(1-\Delta_k)/\pi$. The variation between two functions are not well defined while at least we should note that such value could not be zero since $\phi^+$ and $\tilde{\phi}^+$ are not independent. Expanding them in terms of spherical harmonics, one will see that the variation of the two sources with respect to the basis can be written as
\begin{equation}
    \delta \phi^+(z,\bar{z};k)=\sum_{l\neq 0,m} a^+_{lm}(k)\delta Y^l_m(z,\bar{z})\qquad  \delta \tilde{\phi}^+(z,\bar{z};k)=\sum_{l\neq 0,m} a^-_{lm}(k)\delta Y^l_m(z,\bar{z}) 
\end{equation}
Physically, one can treat the deviation of the basis $\delta Y^l_m$ from the spherical harmonics as the deformation of the background geometry away from purely flat case. For the coefficients $a_{lm}^{\pm}(k)$, they come from the decomposition of the bulk fields $\Phi$ and they are determined by assigning data on the Cauchy hypersurface chosen as the initial time \footnote{Actually, one should further impose Lorentz invariance, causality condition and the cluster decomposition principle on these coefficients when using quantum field theory to calculate scattering amplitudes of particles.}. One can see more discussions on the coefficients in the appendix \ref{Harmonic} or section \ref{Shock}. Therefore one can define the variation between two sources as 
\begin{equation}
    \frac{\delta\phi^+(z,\bar{z};k)}{\delta \tilde{\phi}^+(z',\bar{z}';k)}:= \frac{1}{N_k}\sum_{l\neq 0,m}\frac{a^+_{lm}(k)}{a^-_{lm}(k)}\delta(z-z'),
\end{equation}
in which the factor $N_k=\sum_{l\neq 0,m}$ 1 is introduced for normalization and one can interpret it as the measure of the discrete parameter space $(l,m)$. Following such convention, then we obtain the two-point function
\begin{equation}
    \langle \mathcal{O}(z,\bar{z};k)\mathcal{O}(z',\bar{z}';k) \rangle=\frac{1}{N_k}\sum_{l\neq 0,m}\frac{a^+_{lm}(k)}{a^-_{lm}(k)}\frac{c_k}{|z-z'|^{2\Delta_k}},
\end{equation}
and
\begin{equation}
    \langle \tilde{\mathcal{O}}(z,\bar{z};k)\tilde{\mathcal{O}}(z',\bar{z}';k) \rangle=\frac{1}{N_k}\sum_{l\neq 0,m}\frac{a^-_{lm}(k)}{a^+_{lm}(k)}\frac{c_k}{|z-z'|^{2\Delta_k}}.
\end{equation}
\begin{figure}
    \centering
    \begin{tikzpicture}
\draw(0,1) ellipse (1 and 0.5);
\draw(-1.5,1) node{$\Delta_1$};
\draw(1.5,1) node{$\Delta_2$};
\draw(1.5,-1) node{$\Delta_3$};
\draw(0,-1) ellipse (1 and 0.5);
\filldraw  (-1,1) circle (1.5pt);
\filldraw  (0,-1) circle (1.5pt);
\filldraw  (1,-1) circle (1.5pt);
\draw(-1,1)[->]--(-0.5,0);
\draw(-0.5,0)--(0,-1);
\draw[->](1,1)--(0.5,0);
\draw(0.5,0)--(0,-1);
\draw(1,-1)--(0,-1);
\filldraw  (1,1) circle (1.5pt);
\end{tikzpicture}
\hspace{6.0em}
\begin{tikzpicture}
\draw(0,1) ellipse (1 and 0.5);
\draw(0,-1) ellipse (1 and 0.5);
\draw(-0.5,0.25) node{$\Delta_1$};
\filldraw  (-1,-1) circle (1.5pt);
\filldraw  (0,0.5) circle (1.5pt);
\filldraw  (0,-1) circle (1.5pt);
\draw(-1,-1)--(0,-1);
\draw(1,-1)--(0,-1);
\draw[->](0,0.5)--(0,-0.25);
\draw(0,-0.25)--(0,-1);
\draw(-1.5,-1) node{$\Delta_2$};
\draw(1.5,-1) node{$\Delta_3$};
\filldraw  (1,-1) circle (1.5pt);
\end{tikzpicture}    
    \caption{For $\lambda_3\Phi^3(X)$ interaction, two kinds of $k$-mode contribution for the three point function $\langle\mathcal{O}\mathcal{O}\mathcal{O}\rangle$are shown in the figure. The diagram on the left represents mode contribution like $\phi_1\phi_2\tilde{\phi}_3$ while the mode contribution like $\phi_1\tilde{\phi}_2\tilde{\phi}_3$ are shown on the right. }
\label{3pt+}
\end{figure}
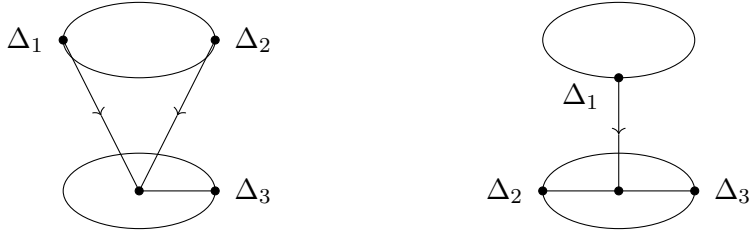
These two kinds of propagators carry the dynamical information of the physical system in the bulk. From the boundary point of view, they describe the coupling of the two series of operators and we represent such relation in Figure \ref{propagators}. For the higher point correlation functions, one needs to take the interaction of particles into consideration. Suppose that we have turned on the $\Phi^3$ interaction of coupling constant $\lambda_3$, then we can write the action as
\begin{eqnarray}
   \lambda_3 \int d^4X\; \Phi^3(X)&=&\lambda_3\int_{AdS_3}d^3x\sqrt{g}\int_\mathcal{P} dk_1dk_2dk_3\int_0^\infty d\tau \frac{1}{\tau} (\tau^{\beta_+^1+\beta_+^2-\beta_+^3+1}\phi(z,\bar{z},\rho;k_1)\\ &&\times\phi(z,\bar{z},\rho;k_2)\tilde{\phi}(z,\bar{z},\rho;k_3)+\tau^{\beta_+^1-\beta_+^2-\beta_+^3+1}\phi(z,\bar{z},\rho;k_1)\tilde{\phi}(z,\bar{z}\rho;k_2)\tilde{\phi}(z,\bar{z}\rho;k_3)+\cdots),\nonumber\label{interaction}
\end{eqnarray}
in which we have decomposed the fields into the integral over $k$-modes and collected all the $\tau$-modes.
To discuss the integral over $\tau$ modes in a more precise way, we write the value of $\beta_+$ as a complex number into the real and imaginary part
\begin{equation}
    \sqrt{1+k_i^2}\equiv\beta_+^i=\gamma_i+ip_i,
\end{equation}
therefore the integral, taking the first one $\phi_1\phi_2\tilde{\phi}_3$ for example, becomes
\begin{equation}
    \int_0^\infty d\tau \frac{1}{\tau}\; \tau^{\gamma_1+\gamma_2-\gamma_3+1} e^{i(p_1+p_2-p_3)\ln{\tau}}\sim  \delta(p_1+p_2-p_3)
\end{equation}
after imposing the condition for the real part
\begin{equation}
    \gamma_1+\gamma_2-\gamma_3=-1.
\end{equation}
The above relation tells us that, in order to describe the $\Phi^3$ interaction, extra modes apart from principal series should be taken into consideration. In such the case, the interacting part in the action then can be reduced to 
\begin{equation}
    \delta(p_1+p_2-p_3)\; \lambda_3\int_{AdS_3}d^3x\sqrt{g}\;\phi(z,\bar{z},\rho;k_1)\phi(z,\bar{z},\rho;k_2)\tilde{\phi}(z,\bar{z},\rho;k_3)
\end{equation}
and its contribution to the three-point function is shown in the left hand side Figure \ref{3pt+}. Moreover, one can see that such three-point functions on the boundary are extremal $\Delta_3=\Delta_1+\Delta_2$ \cite{Lee:1998bxa,DHoker:1999jke} after employing the flat/CFT dictionary \eqref{scale}. For the $\phi_1\tilde{\phi}_2\tilde{\phi}_3$ contribution in \eqref{interaction}, if one imposes the condition 
\begin{equation}
    \gamma_1-\gamma_2-\gamma_3=-1
\end{equation}
therefore the interaction takes the form
\begin{equation}
    \delta(p_1-p_2-p_3)\; \lambda_3\int _{AdS_3}d^3x \sqrt{g} \phi(z,\bar{z},\rho;k_1)\tilde{\phi}(z,\bar{z},\rho;k_2)\tilde{\phi}(z,\bar{z},\rho;k_3)
\end{equation}
and the diagram is shown in the right hand side of Figure \ref{3pt+}. In the figure, we have seen that the internal vertex is inserted in the $-$ disk since we are calculating the three point function $\langle \mathcal{O}\mathcal{O}\mathcal{O}\rangle$ generated by the source $\tilde{\phi}^+$ living on the - disk. For the three point function $\langle\tilde{\mathcal{O}}\tilde{\mathcal{O}}\tilde{\mathcal{O}}\rangle$ we can also show the $\phi\phi\tilde{\phi}$ and $\phi\tilde{\phi}\tilde{\phi}$ interaction in terms of diagrams but now the internal vertex is inserted on the $+$ disk shown in Figure \ref{3pt-}. One can also check such three-point functions are also extremal $\tilde{\Delta}_3=\tilde{\Delta}_1+\tilde{\Delta}_2$ in the sense of shadow operators $\tilde{\Delta}=2-\Delta$.

Following the similar procedure, one can also study the $\Phi^4$ interaction written as 
\begin{equation}
    \lambda_4 \int d^4X \Phi^4(X),
\end{equation}
in which the coupling constant $\lambda_4$ now becomes dimensionless and it is worthwhile to note that, not like the $\Phi^3$ interaction, all the modes on the principal series could contribute to the interaction since the relation
\begin{equation}
    \pm \gamma_1\pm \gamma_2\pm \gamma_3\pm \gamma_4=0
\end{equation}
is satisfied if one sets $\gamma_i=0$, i.e. $\Delta_i=1+ip_i$.
Corresponding diagrams for four-point functions could also been drawn while it is interesting to see that the diagrams introduced for the $\Phi^3$ and $\Phi^4$ interaction can be treated as the intermediate between Feynman and Witten diagrams. If one collapses the two disks in the diagram, i.e. ignoring the dynamical or the causal structure of the system, then we will obtain the Witten diagram which is often illustrated as a single disk. From the other hand, if one tries to sum over all the diagrams of different $k$ modes, then one will recover the Feynman diagrams which enable us to study the scattering amplitudes for particles.
\subsection{Holographic Dictionary for Onshell Scalar Fields}

\qquad In this section we collate the results above and summarise the process for reading off the holographic data corresponding to an onshell scalar field $\Phi(\tau, \rho, z, \bar{z})$. In general the flat/CFT dictionary is given by

\begin{equation}
    {\rm exp}\;\big(i S^{{\rm ren}}(\Phi)\big)=\Big\langle\; {\rm exp}\;\int_{S^2}\int_\mathcal{P} \left (\mathcal{J}_\Delta\;\mathcal{O}_\Delta+\Tilde{\mathcal{J}}_\Delta\Tilde{\mathcal{O}}_\Delta\right)\;\Big\rangle_{CFT}.\label{dicflat}
\end{equation}
To map the data between two sides, we first express the scalar field as a linear superposition of frequency modes, i.e. 
\begin{equation}
\Phi(\tau, \rho, z, \bar{z}) = \int_\mathcal{P} d k \tau^{\beta_+ - 1} \phi(\rho, z, \bar{z}; k) + \int_\mathcal{P} d k \tau^{\beta_- -1} \Tilde{\phi} (\rho,z, \bar{z}; k) \label{split}
\end{equation}  
where following \eqref{spit1} the two classes of modes can be expressed as
\begin{eqnarray}
\phi(\rho, z, \bar{z}; k) &=& \phi(\rho,z,\bar{z}; \beta_+) + \phi(\rho,z,\bar{z}; \beta_-);  \\
\Tilde{\phi}(\rho, z, \bar{z}; k) &=& \tilde{\phi}(\rho,z,\bar{z}; \beta_+) + \tilde{\phi}(\rho,z,\bar{z}; \beta_-). \nonumber 
\end{eqnarray}
These fields have asymptotic expansions 
\begin{eqnarray}
\phi(\rho, z, \bar{z}; k) &=& \rho^{\beta_+ - 1} \phi^+(z,\bar{z}; k) + \rho^{\beta_- - 1} \phi^-(z,\bar{z}; k)  + \cdots    \\
\Tilde{\phi}(\rho, z, \bar{z}; k) &=& \rho^{\beta_+ - 1} \tilde{\phi}^+(z,\bar{z}; k) + \rho^{\beta_- - 1} \tilde{\phi}^-(z,\bar{z}; k) + \cdots \nonumber 
\end{eqnarray}
from which one can read off expectation values and sources according to:
\begin{eqnarray}
\langle {\cal O} (z, \bar{z}; k) \rangle  &=& -2i\beta_+ \phi^- (z, \bar{z}; k) \qquad \mathcal{J}(z,\bar{z};\beta)=\tilde{\phi}^+(z,\bar{z};k)  \label{vev} \\
\langle \tilde{\cal O} (z, \bar{z}; k) \rangle  &=& -2i\beta_+ \tilde{\phi}^- (z, \bar{z}; k) \qquad \tilde{\mathcal{J}}(z,\bar{z};k)={\phi}^+(z,\bar{z};k) \nonumber
\end{eqnarray}
The decomposition of the field \eqref{split} follows from the orthogonality relations:
\begin{eqnarray}
 \phi(\rho, z, \bar{z}; k) &=& \frac{1}{2 \pi} \int_{0}^{\infty} d \tau \; \tau^{\beta_-} \Phi(\tau, \rho, z, \bar{z}) \\
 \Tilde{\phi}(\rho, z, \bar{z}; k) &=& \frac{1}{2 \pi} \int_{0}^{\infty}  d \tau \; \tau^{\beta_+} \Phi(\tau, \rho, z, \bar{z}). \nonumber
\end{eqnarray}

To calculate the two-point function and reduce the data to single AdS surface, we need to check the expression of $\phi(\rho,z,\bar{z};\beta_\pm)$ and $\tilde{\phi}(\tau,\rho,z,\bar{z};\beta_\pm)$ explicitly. Given the AdS modes as the basis, $\phi$, $\tilde{\phi}$ are characterised  by the coefficient $a^+_{lm}(k)$ $a^-_{lm}(k)$ written as
\begin{eqnarray}
    \phi^\pm(\rho,z,\bar{z};k)&=&\sum_{lm} a^+_{lm}(k)\;\phi_l(\rho;\beta_\pm) Y_m^l(z,\bar{z}) \\
      \tilde{\phi}^\pm(\rho,z,\bar{z};k)&=&\sum_{lm} a^-_{lm}(k)\;\phi_l(\rho;\beta_\pm) Y_m^l(z,\bar{z}),
\end{eqnarray}
where we have chosen the normalisation for the spatial function as $\phi_l(\rho;\beta_+)=\rho^{\beta_+-1}+\cdots$. In such case, one can then write the sources in terms of spherical harmonic functions as
\begin{eqnarray}
     \phi^+(z,\bar{z};k)&=&\sum_{l,m} a^+_{lm}(k) Y^l_m(z,\bar{z}) \\   \tilde{\phi}^+(z,\bar{z};k)&=&\sum_{l,m} a^-_{lm}(k) Y^l_m(z,\bar{z}) .
\end{eqnarray}
and the two copies of propagators are given by 
 \begin{eqnarray}
    \langle \mathcal{O}(z,\bar{z};k)\mathcal{O}(z',\bar{z}';k) \rangle&=&\frac{1}{N_k}\sum_{l\neq 0,m}\frac{a^+_{lm}(k)}{a^-_{lm}(k)}\frac{c_k}{|z-z'|^{2\Delta_k}}\\
    \langle \tilde{\mathcal{O}}(z,\bar{z};k)\tilde{\mathcal{O}}(z',\bar{z}';k) \rangle&=&\frac{1}{N_k}\sum_{l\neq 0,m}\frac{a^-_{lm}(k)}{a^+_{lm}(k)}\frac{c_k}{|z-z'|^{2\Delta_k}}.
\end{eqnarray}
Such results work for Minkowski spacetime. For asymptotically Minkowski spacetime, one needs to do the harmonics analysis and the results are shown in Appendix \ref{Harmonic} as \eqref{2pth1} and \eqref{2pth2}. 

Moreover, for higher point functions and theories with interaction. It is convenient to represent the correlation functions using two copies of disks labeled by + and -. For the correlators constructed out of the operator $\mathcal{O}$, the interactions are described by the internal vertices inserted on the - disk while for the $\tilde{\mathcal{O}}$ correlators, the points will be inserted on the + disk, i.e, all the internal vertices of each diagram can only exist in one of the disk. For the external legs, the one connects two points on a single disk is described by the standard AdS/CFT propagator. While for the ones that connect two disks, we need to take the extra factors constructed out of the coefficients $a^{\pm}_{lm}(k)$ into consideration. If we assume that the legs between two disks have directions and they always flow into the internal points, the leg that starts from the $i$th external vertex on the + disk then ends at the internal point the on - disk will contribute to a factor
\begin{equation}
    i(+) \longrightarrow  \bullet (-) \qquad \qquad \frac{1}{N_k}\sum_{l\neq 0,m}\frac{a^+_{lm}(k_i)}{a^-_{lm}(k_i)}
\end{equation}
and the one  starts from the $i$th external vertex on the - disk then ends at the internal point the on + disk will contribute to a factor
\begin{equation}
    i(-)\longrightarrow \bullet(+) \qquad \qquad \frac{1}{N_k}\sum_{l\neq 0,m}\frac{a^-_{lm}(k_i)}{a^+_{lm}(k_i)}.
\end{equation}
\section{Shock Waves and Their Holographic Interpretation}\label{Shock}
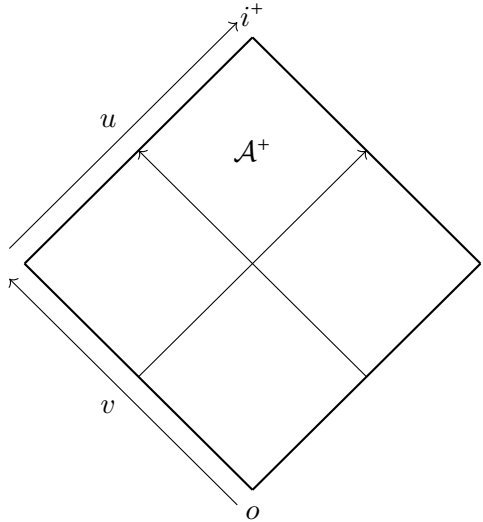
\begin{figure}[h!]
    \centering
    \begin{tikzpicture}
\draw[thick](0,3)--(3,0);
\draw[thick](3,0)--(0,-3);
\draw[thick](-3,0)--(0,-3);
\draw[thick](-3,0)--(0,3);
\draw[->](-1.5,-1.5)--(1.5,1.5);
\draw[<-](-1.5,1.5)--(1.5,-1.5);
\draw (0,3.3) node{$i^+$};
\draw (0,-3.3) node{$o$};
\draw (0,+1.5) node{$\mathcal{A}^+$};
\draw[->](-0.2,-3.2)--(-3.2,-0.2);
\draw(-1.9,-1.9) node{$v$};
\draw[<-](-0.2,3.2)--(-3.2,0.2);
\draw(-1.9,1.9) node{$u$};
\end{tikzpicture}    
    \caption{In and out going shock waves propagating in region $\mathcal{A}^+$ are shown in the figure.}
    \label{inout}
\end{figure}
In this section we consider the holographic interpretation of a shock wave. Here in our context, the shock waves are scalar shocks that could either distribute on a spherical shell or localise along the null geodesic of a massless particle. For the spherical shock wave, it describes the wave caused by a point like source then propagates in spacetime following a homogeneous way. For the second kind of shock wave, it could be treated as the approximation for the signals traveling at the speed of light like the laser beam \footnote{We assume such method can be generalised to gauge fields or we are dealing with a high energy beam of bosonic particles with small mass. Maybe for massive particles, one should consider the ingoing and outgoing wavepackets.}. In fact, to make photons trapped in the beam, one should take gravity effects into consideration and it turns out such shock wave will induce backreactions on the metric studied in \cite{Aichelburg:1970dh,Dray:1984ha}. Then the massive case is also studied in a perturbative way. However, in our situation, the shape of shock waves will be less important while the ingoing and outgoing behaviour of the wave will be crucial. To construct the shock wave solutions, we start from the Minkowski metric  written as
\begin{eqnarray}
    ds^2&=&-dt^2+dr^2+r^2\;d\Omega^2_2 \nonumber \\
        &=&-dudv+r^2\;d\Omega^2_2, \label{standard}
\end{eqnarray}
where $t$, $r$ are the time and radial directions and $d\Omega^2_2$ is the standard 2 sphere metric. In the second line, the retarded and advanced coordinates $u$, $v$ are defined as
\begin{equation}
    u=t-r,\qquad v=t+r.
\end{equation}
A massless particle which is described by the field $\Phi$ satisfies 
\begin{equation}
    -4\partial_u\partial_v \Phi+ \frac{1}{r^2}\square_{S^2}\Phi=0.
\end{equation}
Let us consider a spherically symmetric solution so that the equation reduces to 
\begin{equation}
    \partial_u\partial_v \Phi=0,
\end{equation}
and general solutions are given by 
\begin{equation}
    \Phi(u,v)=\phi(u)+\tilde{\phi}(v)\label{solution}
\end{equation}
where $\phi$ and $\tilde{\phi}$ are arbitrary functions of $u$, $v$. 

An interesting physical solution is the spherical shock wave. A shock wave emitted from the boundary and propagating along the null ray as illustrated in Figure \ref{inout} is described by $\Phi^{in}_s$
\begin{equation}
    \Phi^{in}_s(v)=\phi_0 \;\delta(v-v_0), \qquad v_0>0.
\end{equation}
We can view the shock wave solution as a specific linear combination of plane wave solutions. Furthermore, to make the wave really localise along the null ray, one could consider the gravitational  shock wave $\Phi_g^{in}=\phi_0\delta(v-v_0)\delta(z-z_0)$ while it is not a solution for KG equation in flat spacetime and it only exists when the gravitational effect is taken into consideration. Here, to study the flat/CFT dictionary in a simple way, we choose to use the spherical shock wave as an example to perform the calculation and the results for $\Phi^{in}_s$ is obtained by inserting the factor $\delta(z-z_0)$ behind \footnote{Here we assume that the localised shock waves propagate on the background where the holography principle still works. }.

To express the shock wave in terms of modes adapted to the hyperbolic slicing, we need to transform the coordinates \eqref{standard} into Milne coordinates using
\begin{equation}
    t^2-r^2=\tau^2,\qquad \rho\tau=r,
\end{equation}
 in which we should note that the Milne coordinates will only cover the region $\mathcal{A}^+$ if both $(\tau,\rho)$ are required to be positive $\rho,\tau\geq 0$. The near light cone region is described by $\tau \rightarrow 0$ and the asymptotic region is given by $\tau\rightarrow \infty$. In Milne coordinates, the shock wave can be expressed as
\begin{equation}
    \Phi^{in}_s=\phi_0\;\delta(\rho\tau+\tau\sqrt{1+\rho^2}-v_0)=\phi_0\delta(\tau e^\eta-v_0).
\end{equation}

We can now decompose this solution into modes as described in the previous section, resulting in 
\begin{eqnarray}
\Phi^{in}(\rho, z, \bar{z}; \beta_+) &=& \frac{\phi_0}{2 \pi} v_0^{\beta_-} e^{- (1+ \beta_-)\eta},  \\
\Phi^{in}(\rho, z, \bar{z}; \beta_-) &=& \frac{\phi_0}{2 \pi}  v_0^{\beta_+} e^{- (1+ \beta_+)\eta}. \nonumber
\end{eqnarray}
The fields are independent of the sphere coordinates. One can immediately read off the coefficients of the asymptotic expansion using the relation $\rho = \sinh \eta$ as
\begin{eqnarray}
\phi^+ (z, \bar{z}; k) &=&  \frac{\phi_0}{2 \pi} 2^{\beta_+  - 1} v_0^{\beta_-} \qquad \phi^-(z, \bar{z}; k) = 0; \\
\tilde{\phi}^+ (z, \bar{z}; k) &=&  0 \qquad \tilde{\phi}^-(z, \bar{z}; k) = \frac{\phi_0}{2 \pi} 2^{- \Delta} v_0^{\beta_+}. \nonumber 
\end{eqnarray} 
This means that the operators ${\cal O} (z, \bar{z}; k)$ has no source or expectation value, but the operators ${\tilde{\cal O}}(z,\bar{z};k)$ have both: the sources are 
$\phi^+(z,\bar{z};k)$ while
\begin{equation}
\langle \tilde{O} (z, \bar{z}; k) \rangle = -i\beta_+ \frac{\phi_0}{ \pi} 2^{- \Delta} v_0^{\beta_+} .
\end{equation}

It is straightforward to repeat the same exercise for a shock wave propagating along an orthogonal null ray i.e.
\begin{equation}
    \Phi^{out}_s(u)=\phi_0 \;\delta(u - u_0), \qquad u_0>0.
\end{equation}
This can be expressed in terms of the Milne coordinates as 
\begin{equation}
    \Phi^{out}_s=\phi_0\;\delta(\tau\sqrt{1+\rho^2} - \tau \rho - u_0)=\phi_0\delta(\tau e^{-\eta} - u_0).
\end{equation}
Decomposing into modes one finds
\begin{eqnarray}
\Phi^{out}(\rho, z, \bar{z}; \beta_+) &=& \frac{\phi_0}{2 \pi} u_0^{\beta_-} e^{ (1+ \beta_-) \eta},  \\
\Phi^{out} (\rho, z, \bar{z}; \beta_-) &=& \frac{\phi_0}{2 \pi}  u_0^{\beta_+} e^{ (1+ \beta_+)\eta} .\nonumber
\end{eqnarray}
One can then read off the coefficients of the asymptotic expansion using the relation $\rho = \sinh \eta$ as
\begin{eqnarray}
{\phi}^+ (z, \bar{z}; k) &=&  0 \qquad {\phi}^-(z, \bar{z}; k) = \frac{\phi_0}{2 \pi} 2^{\Delta} u_0^{\beta_+}; \\
\tilde{\phi} ^+ (z, \bar{z}; k) &=&  \frac{\phi_0}{2 \pi} 2^{1  - \beta_+} u_0^{\beta_-} \qquad \tilde{\phi}^-(z, \bar{z}; k) = 0. \nonumber
\end{eqnarray} 
This means that the operators $\tilde{\cal O} (z, \bar{z}; k)$ has no source or expectation value, but the operators ${\cal O}(z,\bar{z};k)$ have both: the sources are 
$\tilde{\phi}^+(z,\bar{z};k)$ while
\begin{equation}
\langle {O} (z, \bar{z}; k) \rangle =-i \beta_+ \frac{\phi_0}{ \pi} 2^{\Delta} u_0^{\beta_+}.
\end{equation}

Thus we can understand the two sets of dual operators as describing modes propagating in $(u,v)$ directions respectively:
\begin{eqnarray}
&& \Phi(u) \rightarrow \{ {\cal O} (z,\bar{z};k), \tilde{\phi}^+(z,\bar{z};k) \}; \\
&& \Phi(v) \rightarrow \{ \tilde{\cal O} (z, \bar{z};k), \phi^+(z, \bar{z}; k) \}. \nonumber
\end{eqnarray}
Based on above results, we see that the operators introduced in this article have the same physical interpretation as the operators studied by the celestial holography and they should be equivalent from kinematic point of view since they carry the same amount of data while a careful analysis shows that they are not exactly the same. We are extracting the data of bulk fields along the AdS hyperboloid by sending $\rho$ to infinity. During such procedure, all the operators will fall into the same sphere localised at the boundary of light cone with different angles labelled by $\tau$ therefore it is more reasonable to make our CFT operators live on the single sphere. For the celestial holography approach, the bulk data is extracted by sending the field to the two null boundaries following the null direction that $u$ or $v$ is constant so celestial operators live on family of spheres on the null boundary. These two ways of decomposing bulk data should be equivalent and related by the coordinate transformation between Milne coordinates and advanced/retarded coordinates. \footnote{It seems like the celestial ingoing operator is related to the operator $\tilde{\mathcal{O}}$ at the boundary of $\mathcal{A}^-$ while the celestial outgoing operator is related to the operator $\mathcal{O}$ at the boundary of $\mathcal{A}^+$. Following two kinds of definitions, the corresponding two-point functions are different while we find that the new defined celestial dictionary \cite{Pasterski:2017kqt,Crawley:2021ivb,furugori2023celestial} will produce the same two-point function as ours which are the standard CFT correlation functions. It is interesting to explore the relation between our operators $\mathcal{O},\tilde{\mathcal{O}}$ and the new defined operator $\mathcal{O}^+,\mathcal{O}^-$ further where $\mathcal{O}^+$ is the shadow of the standard ingoing celestial operators and $\mathcal{O}^-$ is the outing celestial operators.}

As for the two point functions, the structure will become complicated and one needs to take the gravitational effect into consideration. For the spherical shock wave, it is the solution for KG equation in Minkowski but the two-point function will become trivial since the solution takes constant value on the sphere and the method we have introduced in the section \ref{Dictionary} will not work. It does not mean that the dual theory on the boundary will become trivial while we need to take the  gravity backreaction into consideration in order to investigate the correlation function at higher order if one treats the constant $\phi_0$ as a small parameter. After backreaction from the matter, the metric then becomes
\begin{equation}
    G'_{\mu\nu}=G_{\mu\nu}+\delta G_{\mu\nu},
\end{equation}
which $G$ is the metric for Minkowski and the deformation caused by the matter is denoted as $\delta G$. They are governed by Einstein equation  
\begin{equation}
    R_{\mu\nu}-\frac{1}{2}R\;G'_{\mu\nu}=T_{\mu\nu},
\end{equation}
in which $R_{\mu\nu}$ is the Ricci curvature for $G'$ and $T_{\mu\nu}$ is the stress tensor determined by the scalar profile and in our case it is the shock wave $\Phi_s$ thus we can see that the stress tensor is of the order $\phi_0^2$. One can treat it as the Newtonian  constant $\phi^2_0\sim G_N$ which is not explicitly shown in the equation.  From the above equation, one can also see that the deformation $\delta G$ also goes as the order of $\phi_0^2$ \footnote{In fact we have $\langle T^{\rm CFT}_{\mu\nu}\rangle\sim \delta G$ in which $T^{\rm CFT}_{\mu\nu} $ is the stress tensor of the dual CFT theory on the celestial sphere. The specific expression relies on the holographic renormalisation of Einstein-Hilbert action which has been done in Graham-Fefferman coordinates for AdS case \cite{deHaro:2000vlm}. }. Given the deformed background then the scalar fluctuation $\delta \Phi$ on the shock wave profile is determined by the KG equation
\begin{equation}
    \square_{G'}\Phi'=0,
\end{equation}
where $\Phi'=\Phi_s+\delta\Phi$. Here we should note that, although $\Phi_s$ is constant on the celestial sphere while the fluctuation $\delta \Phi$ is not necessary constant and it depends on the further specification of the data at the initial time. Therefore both of the vacuum expectation value and source will receive correction of order $\phi^2_0$ coming from $\delta \Phi$ and the two-point function now becomes
\begin{equation}
    \langle \mathcal{O}(z;k)\mathcal{O}(z';k)\rangle_s=\langle \mathcal{O}(z;k)\mathcal{O}(z';k)\rangle  +\phi_0^2\;F(z,z';k),
\end{equation}
where $\langle\cdots\rangle_s$ represents that the operators are now inserted on the shock wave background rather than the Minkowski vacuum $\langle\cdots \rangle$ and the higher order correction is of the order $\phi_0^2$, i.e. $G_N$.  Its specific form is given by the function $F(z,z';k)$ determined by the variation $\delta \Phi$. From above discussion, we know that all the spherical solutions without considering gravity effect in Minkowski are degenerate from boundary point of view and one needs to consider the variation of the scalar field in order to distinguish all the spherical solutions. The broken of the spherical symmetry caused by the gravity effect will enable us to calculate two-point functions at leading order and then introduce subleading terms characterised by the function $F(z,z';k)$. For the localised shock wave, one needs to figure out the background and then check if the holography principle still works on such background, which depends on the definition of asymptotic flat as well as the ability of holography principle and such work goes beyond the scope of this article.

\subsection{Coefficients}
\qquad  After the study of the dual correlation functions on the boundary. Here we will use the shock wave model as an example to study the bulk field in a direct way following the mode analysis introduced in section \ref{Mode} and try to determine the coefficients of those modes. It is easier for the spherical shock waves since they are constant on the sphere and only the zero mode will contribute when performing the mode expansion. The analysis for localised shock wave will be harder since the analysis for the $l\ge 1$ mode will be difficult and we will leave the mode analysis for $\Phi_g$ for further investigation.\\

{\it Massless Fields}
\bigskip

From the discussion in section \ref{radialeq} and appendix \ref{solutions}, we have seen that the zero mode $l=0$ on the AdS hyperboloid has two independent solutions at large radius $\rho=\sinh{\eta}\rightarrow \infty$
 \begin{equation}
     \phi_0(\eta;\beta_+)=\frac{e^{\beta_+\eta}}{\sinh{\eta}},\qquad \phi_0(\eta;\beta_-)=\frac{e^{\beta_-\eta}}{\sinh{\eta}}.
 \end{equation}
 The regular solution at $\rho=0$, denoted as $\phi_r(\eta;k)$, is the linear combination of them with ratio $C^-_0(k)/C^+_0(k)=-1$ \footnote{One can obtain this by the direct observation of the liner combination of $\phi_0(\eta;\beta_\pm)$ or by checking the formula of $C^{\pm}_0(k)$ for odd $\beta$ in the Appendix \ref{solutions}.} thus it can be written as    
\begin{equation}
    \phi_r(\eta;k)=\frac{1}{\sqrt{\pi}}\;\frac{\sinh{\beta_+\eta}}{\sinh{\eta}}.
\end{equation}
One can check that $\phi_r$ is regular for arbitrary $\beta_-$ since $\phi_r\sim \beta_+$ at $\eta=0$. Here we are interested in the principal series case $\beta_+=ik$ for $k\geq 0$ and we assume that the result for other value can be obtained by the analytic continuation of $\beta_+$.

For the ingoing waves, one has the expansion
\begin{equation}
    \Phi^{in}(v)=\int_\mathcal{P} dk \;(a^+_{in}(k)\; \tau^{-1+\beta_+}+a^-_{in}(k)\;\tau^{-1+\beta_-})\;\phi_r(\eta,k),
\end{equation}
in which  $a^\pm_{in}(k)$ is the pair of coefficients that we are going to determine. To calculate these coefficients, one should first note the orthogonal relation
\begin{equation}
   \int_{-\infty}^{+\infty} \sinh^2{\eta} \;\phi^*_r(\eta;k)\phi_r(\eta;k)=\delta(k-k'),
\end{equation}
in which $\phi_r^*$ is the complex conjugate of $\phi_r$. Given the above relation, one can project out the $\eta$ dependent part by performing the integral
\begin{equation}
\phi_0 \int_{-\infty}^{+\infty} d\eta\; \delta(\tau e^{\eta}-v_0) \sinh^2\eta\; \phi^*_r(\eta;k)
\end{equation}
therefore coefficients $a^\pm(k)$ are then deduced to be 
\begin{equation}
    a^+_{in}(k)=\frac{\phi_0 v_0^{-\beta_+}}{4\sqrt{\pi}},\qquad a^-_{in}(k)=-\frac{\phi_0v_0^{\beta_+}}{4\sqrt{\pi}},
\end{equation}
in which we have omitted the correction term of order $\tau^2$. The fact that we get extra terms in additional to the modes $\tau^{-1+\beta_{\pm}}$ implies that the basis we have chosen is not complete. Here we assume that the mode expansion is done near the Milne horizon thus $\tau\rightarrow0$ and the higher order term will be subleading.
For the outgoing shock wave, following similar procedure, one has the expansion 
\begin{equation}
    \Phi^{out}(u)=\int_{\mathcal{P}} dk \;(a^+_{out}(k)\; \tau^{-1+\beta_+}+a^-_{out}(k)\;\tau^{-1+\beta_-})\;\phi_r(\eta,k),
\end{equation}
in which the corresponding coefficients $a^{\pm}_{out}(k)$ are determined to be
\begin{equation}
    a^+_{out}(k)=\frac{\phi_0 u_0^{\beta_+}}{4\sqrt{\pi}},\qquad a^-_{out}(k)=-\frac{\phi_0u_0^{-\beta_+}}{4\sqrt{\pi}}.
\end{equation}
\\\\
{\it Massive Fields}
\bigskip

Now we turn to the study of massive particles. First we try to make the particle slightly massive and then investigate the perturbative behaviour of the solution around the spherical shock wave. Similar to the study of massive KG equation, we choose to write the equation of motion for massive particle as
\begin{equation}
(\partial_u\partial_v +\lambda M^2)\Phi_M(X)=0,
\end{equation}
in which $M$ is a constant and $\lambda$ is a small parameter that represents the mass of the particles is small. Then we can write a general solution for the massive equation up to the first order of $\lambda$ as
\begin{equation}
    \Phi_M(X)=\phi_0\delta(v-v_0)+\lambda f(u,v),
\end{equation}
in which $f(u,v)$ is a function of $u,v$. To determine $f(u,v)$, one should substitute the solution into the equation and solve the equation by the order of $\lambda$ then obtain
\begin{equation}
    \Phi_M(X)=\phi_0\delta(v-v_0)-u\;\lambda\;\phi_0 M^2\;\theta(v-v_0), \label{per}
\end{equation}
in which $\theta(v-v_0)$ is the step function supported in the region $v>v_0$. The step function correction term tells us that, by adding a small amount of mass, the shock wave will be no longer localised along some spherical shell and propagate along the null direction while it will have a tail spreading over the whole region $v>v_0$. If we define the coefficients of massive field expanded by the modes on AdS surfaces as $a_M^\pm(k)$, i.e.
\begin{equation}
    \Phi_M(u,v)=\int_\mathcal{P} dk \;(a^+_{M}(k)\; \tau^{-1+\beta_+}+a^-_{M}(k)\;\tau^{-1+\beta_-})\;\phi_r(\eta,k).
\end{equation}
Then one can write $a_M^\pm(k)$ by the order $\lambda$ as
\begin{equation}
    a_{M}^{\pm}(k)=a_{0}^{\pm}(k)+\lambda\; a_{1}^{\pm}(k)+\lambda^2 a^{\pm}_{2}(k)+\cdots,
\end{equation}
in which $a^{\pm}_{0}(k)$ are the coefficients for massless case we have discussed before
\begin{equation}
    a^{\pm}_0(k)=a^{\pm}_{in}(k)
\end{equation}
and $a^{\pm}_{i}(k)$ are higher order terms. Taking the solution in \eqref{per} for example, to calculate $a_1^\pm(k)$ one should evaluate the integral
\begin{equation}
    \int_{-\infty}^{+\infty}d\eta \;\sinh^2{\eta}\;e^{-\eta}\tau\;\phi_r^*(\eta;k)\theta(\tau e^{\eta}-v_0),
\end{equation}
in which we still use the massless solution as the basis when performing the perturbative expansion. The above integral will vanish when $\tau$ goes to zero thus one can conclude that
\begin{equation}
    a^{\pm}_1(k)=0,
\end{equation}
which tells us coefficients are stable at the massless case. It shows that, for the modes we are interested in, the mass of particle will not play a crucial role and make significant contribution thus the shock wave model is till good approximation for particles with small mass.
\subsection{Cauchy Problem and Scattering}

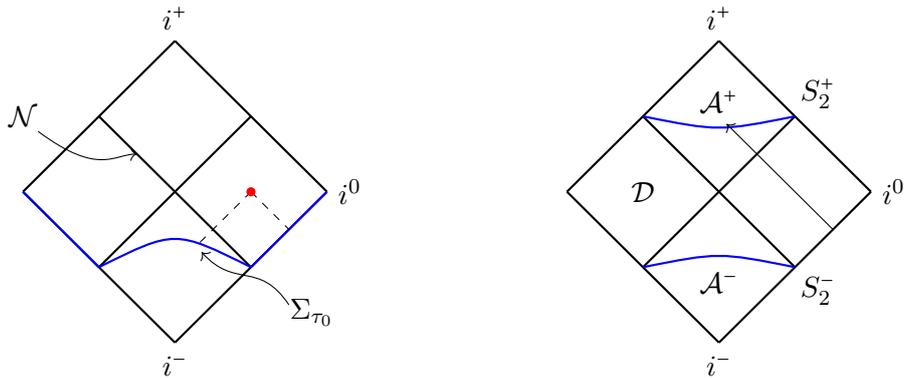
\begin{figure}[h!]
    \centering
    \begin{tikzpicture}
\draw[thick](0,2)--(2,0);
\draw[thick](2,0)--(0,-2);
\draw[thick](-2,0)--(0,-2);
\draw[thick](-2,0)--(0,2);
\draw[thick](-1,-1)--(1,1);
\draw[thick](-1,1)--(1,-1);
\draw[thick,blue](-1,-1)..controls(0,-0.5)..(1,-1);
\draw[thick,blue](-2,0)--(-1,-1);
\draw[thick,blue](2,0)--(1,-1);
\draw (2.3,0) node{$i^0$};
\draw[dashed](1,0)--(1.5,-0.5);
\draw[dashed](1,0)--(0.3,-0.7);
\draw (0,2.3) node{$i^+$};
\draw (0,-2.3) node{$i^-$};
\filldraw [red] (1,0) circle (1.5pt);
\draw[<-] (0.35,-0.8) .. controls (0.8,-1.3)and(1.1,-0.9) .. (1.5,-1.5);
\draw[<-] (-0.5,0.5) .. controls (-0.8,0.5)and(-1,0.2) .. (-1.8,0.8);
\draw(-2,1) node{$\mathcal{N}$};
\draw(1.8,-1.6) node{$\Sigma_{\tau_0}$};
\end{tikzpicture}
\hspace{6.0em}
\begin{tikzpicture}
\draw[thick](0,2)--(2,0);
\draw[thick](2,0)--(0,-2);
\draw[thick](-2,0)--(0,-2);
\draw[thick](-2,0)--(0,2);
\draw[thick](-1,-1)--(1,1);
\draw[thick](-1,1)--(1,-1);
\draw[thick,blue](-1,-1)..controls(0,-0.8)..(1,-1);
\draw[thick,blue](-1,1)..controls(0,0.8)..(1,1);
\draw (2.3,0) node{$i^0$};
\draw(1.3,1.3) node{$S^+_2$};
\draw(1.3,-1.3) node{$S^-_2$};
\draw (0,2.3) node{$i^+$};
\draw (0,-2.3) node{$i^-$};
\draw(0,1.2) node{$\mathcal{A^+}$};
\draw(0,-1.2) node{$\mathcal{A^-}$};
\draw[->](1.5,-0.5)--(0.1,0.9);
\draw(-1,0) node{$\mathcal{D}$};
\end{tikzpicture}    
    \caption{A single AdS surface together with part of null boundary form a Cauchy surface for the whole Minkowski spacetime shown the left hand side. For example, to determine the field configuration at the red point, one needs to specify the data on both of the AdS hyperboloid and null boundary. The shock wave will transform the data from the null boundary to the other AdS surface in region $\mathcal{A}^+$ so that two copies of AdS surfaces are equivalent to a Cauchy surface, which is illustrated in the right figure. }
    \label{cauchy}
\end{figure}
In section \ref{Dictionary}, we start from the holographic renormalisation for the onshell action in region $\mathcal{A}^+$ then conclude that the theory in flat spacetime $\mathcal{A}^+$ is dual to the CFT on the celestial sphere $S_2^+$ located at the future null boundary. To study the whole Minkowski space, in principle, one should consider the action in the region $\mathcal{A}^+\cup \mathcal{D}\cup \mathcal{A}^-$ while it was conjectured in the work \cite{deBoer:2003vf} that it is enough to fully reconstruct the bulk theory using the holographic CFT data on the celestial sphere $S_2^+$ and $S_2^-$ therefore all the information of Minkowski could be classified by specifying the data on the two copies of AdS hyperboloid in $\mathcal{A}^+$ and $\mathcal{A
}^-$ . In particular, the scattering amplitudes in Minkowski can also be constructed by studying the states on these two AdS hyperboloid although we know the fact that they are not the standard Cauchy surface. Here based on the study of AdS and dS modes, we will reconsider the distribution of information in Minkowski and a physical proof of the above conjecture will also be illustrated by doing a thought experiment on the shock wave model.

Following the principle of the mode expansion, to study the local behaviour of the solution $\Phi_M(X)$ in region $\mathcal{A}$ denoted as $\Phi^A_M(X)$, one can expand the solution in terms of modes propagating on the AdS slicing. As we have studied in the section \ref{Mode}, the solution $\Phi_M^A$ can be represented by the linear combination of modes with effective mass $k$ provided that there is a set $\mathcal{P}$ of $k$ in which all the modes together form a complete basis of the solution space, written as
\begin{equation}
    \Phi_M^A(X)=\sum_l\int_{\mathcal{P}_A} dk\; a_{l}(k) \; \psi^A(\tau;k) F^A_{kl}(\rho,z,\bar{z}),
\end{equation}
in which $a_{l}(k)$ are coefficients and the label $l$ is used to represent the other internal variables. $F^A_{kl}(\rho,z,\bar{z})=\phi_{l}(\rho;k)Y^l_m(z,\bar{z})$ are the spatial modes introduced before while $(\rho,z,\bar{z})$ is the coordinate of AdS hyperboloid. For the same reason, we can choose to decompose the solution in region $\mathcal{D}$, denoted as $\Phi_M^{D}(X)$, into $dS$ modes $\psi^D(\rho;k)F_{kl}^D(\tau,z,\bar{z})$ thus it can be written as
\begin{equation}
    \Phi_M^{D}(X)=\sum_{l}\int_{\mathcal{P}_D} dk\; b_{l}(k)\; \psi^D(\rho;k)F_{kl}^D(\tau,z,\bar{z}),
\end{equation}
in which we should note that the position of variable $\tau$ and $\rho$ are switched since we are using them to label the timelike and spacelike direction.

Before imposing the initial condition of the solution $\Phi_M(X)$, we first consider the analytic continuation of the field $\Phi_M(X)$ from the region $\mathcal{A}^-$ into the region $\mathcal{D}$ via the null surface $\mathcal{N}$ shown in Figure \ref{cauchy}. Given the field configuration $\Phi^{A}_M(X)$, one can perform the analytic continuation by making $k\rightarrow ik$ across the null surface $\mathcal{N}$ then obtain $\Phi^{D}_M(X)$. In terms of the coefficients, that is to say
\begin{equation}
 \{ a_{l}(k)\}= \{b_{l}(k)\}
\end{equation}
in which we use the notion $\{\}$ to represent the information contained in the modes and the equal sign means that one can determine all the $b_{l}(k)$ s given the set of $a_{l}(k)$ or vice versa.

To study the initial condition, or to determine the coefficients, first we need to choose a proper codimension one surface to set up the initial data. For the field $\Phi^{A}_M(X)$, one can choose the AdS slicing $X^2=-\tau_0^2$, denoted by $\Sigma_{\tau_0}$ as the Cauchy surface for region $\mathcal{A}^-$ therefore the field in region $\mathcal{A}^-$ is uniquely determined given the initial data $f_i$, $g_i$
\begin{equation}
    \Phi_M(\tau_0,\rho,z,\bar{z})=f_i(\rho,z,\bar{z}),\qquad n^i\partial_i\Phi_M= g_i(\rho,z,\bar{z}) 
\end{equation}
where $n^i$ denotes the further unit normal of $\Sigma_{\tau_0}$. For the field in region $\mathcal{D}$, the data on the surface $\Sigma_{\tau_0}$ is not enough for us to uniquely fix the field configuration $\Phi^{D}_M(X)$. One also needs to specify the data along the null boundary so that they form the Cauchy surface of the whole Minkowski together with the surface $\Sigma_{\tau_0}$, which means one needs more data to determine $\Phi^{D}_M$ comparing to $\Phi^A_M(X)$. Since we have already known that fields in the region $\mathcal{A}$, $\mathcal{D}$ are fully determined by $\{a_{l}(k)\}$ and $\{b_{l}(k)\}$, we conclude that
\begin{equation}
    \{a_{l}(k)\}\subset \{b_{l}(k)\},
\end{equation}
where the symbol $\subset$ means that one can determine all the coefficients $a_l(k)$ given the set of $b_l(k)$ while the other direction is not true anymore, which implies that there are modes not governed by the analytic continuation thus one has $\mathcal{P}_A\subset \mathcal{P}_D$.

Furthermore, based on the calculation in the previous section, we see that, for the massless particle, one can construct the shock wave as the solution of the Klein-Gordon equation. The shock waves propagate along the null direction and they are localised around the trajectory of the massless particles. Moreover, these shock waves that start from the null infinity then go through the AdS slicing surfaces in region $\mathcal{A}^+$ enable the exchange of information between the observer living in some particular AdS surface in region $\mathcal{A^+}$ and the observer on the null boundary. For example, the observer at the boundary can send the information of the initial position and momentum of the particle to the observer in region $\mathcal{A}^+$ via the shock wave and the observer in region $\mathcal{A}^+$ can read out these information by determining the coefficients $a^\pm(k)$. Thus two copies of AdS surface in region $\mathcal{A}^-$ and $\mathcal{A}^+$, respectively, form a  structure that is equivalent to the Cauchy surface since we know that one AdS Surface in region $\mathcal{A}^-$ together with half of the null boundary $v>0$ carry a complete set of data for one to determine the field configuration in the whole space time.

\section{Discussion and Conclusions}

\qquad In this article, we have used the AdS/CFT dictionary to develop a holographic dictionary between flat space and celestial CFT. The key steps in our approach are transforming bulk fields from time to frequency representation, and using the usual AdS/CFT dictionary on spatial hyperbolic slices of the fields in mixed representation of frequency/hyperbolic spatial coordinates. We have shown that a single scalar field propagating in Minkowski is dual to two series of operators on the celestial sphere with scale dimensions on the principal series. One can physically interpret the two sets of operators as ingoing and outgoing modes. 

 Here we have focussed on the example of a scalar field but one would expect an analogous structure for the bulk metric. In particular, one would begin the construction of the holographic dictionary by expressing the 4d metric in a $(3+1)$ form, respecting covariance of the spatial slices and again transforming time dependence into frequency dependence. Working to linear order in the metric perturbations around a fixed background, one would thus obtain fields of the form $\{ h_{ab}(k), h_a(k), h(k) \}$ i.e. spin two, spin one and spin zero from the perspective of the three-dimensional spatial slices. The corresponding dual operators would then be expected to have the structure $\{X_{ij}^{\pm} (k), X_i^{\pm}(k), X^{\pm}(k) \}$, towers of spin two, spin one and spin zero operators. It would be interesting to work out the detailed dictionary in future work.  The renormalisation procedure for the gravity action is expected to be subtle but it will enable determination of central charges as well as facilitate the study of non-trivial gravitational backgrounds such as gravitational shock waves and black holes\footnote{We would like to thank K$\rm \acute{e}$vin Nguyen for pointing out the work on the study of renormalised effective gravitational action on the celestial sphere obtained by reducing the dimension following the dS hyperboloid in the Rindler wedge \cite{,Nguyen:2020hot,Nguyen:2021ydb}. We also found out the discussion of effective gravity action and the central charge in the context of wedge holography \cite{Ogawa:2022fhy}.}.

%To do the flat holography for gauge fields, one also needs to choose a special gauge for the $U(1)$ gauge fields and the reduction of $4d$ Maxwell to $3d$ Chern-Simon should be well explored. 

%Therefore, the answer for the structure $X$ now becomes

%\begin{equation}
 %   X= {\rm Two\; \;series\;\; of \;\;CFT\;\;Operators.}
%\end{equation}

Asymptotically (locally) flat spacetimes have as asymptotic symmetries the (extended) BMS groups at the null boundaries. Therefore the total symmetry for given observable quantities should be ${\rm BMS}^+\times{\rm BMS}^- $ since we have the null boundaries at far past and far future and we proposed that such symmetry is manifested by these two series of operators. Moreover in the work \cite{Strominger:2013jfa}, Strominger proposed that the symmetry which a quantum gravity scattering matrix should preserve is the subgroup ${\rm BMS}^0\subset {\rm BMS}^+\times{\rm BMS}^- $ by matching two null boundaries at the spatial infinity $i^0$. This fits with our observation that the two series operators are dual to the ingoing and outgoing shock waves in the bulk and they are related by physical process that occur in the center. From the boundary point of view, we can see that these two series of operators are coupled with each other.

There has recently been considerable discussion of the role of Carollian symmetry in flat space holography \cite{Hartong:2015usd,Ravera:2019ize,Bagchi:2019xfx,Bergshoeff:2020xhv,Hansen:2021fxi,donnay2022carrollian,Bagchi:2022emh,deBoer:2023fnj,Bagchi:2023ysc,Bagchi:2023fbj,Nguyen:2023vfz,Saha:2023abr,Saha:2023hsl}. It would be interesting to explore how the structure of the holographic dictionary for the metric can be interpreted in term of Carollian structure. In the context of Carollian CFTs, one can introduce the notion of Carrollian time $t_c$ as the dual of effective mass $t_c \sim k$ and thus the series of correlation functions on the celestial sphere can be viewed as dual to a 3d correlation function, i.e.
\begin{equation}
    \langle \mathcal{O}(z,\bar{z};k)\mathcal{O}(z',\bar{z}';k)\rangle \longleftrightarrow     \langle \mathcal{O}(z,\bar{z},t_c)\mathcal{O}(z',\bar{z}',t_c)\rangle,
\end{equation}
and these are related by the integral transform 
\begin{equation}
    \mathcal{O}(t_c,z,\bar{z})=\int_\mathcal{P}dk\; G(t_c,k)\;\mathcal{O}(z,\bar{z};k),
\end{equation}
in which the Green function $G(t_c,k)$ would be determined by the definition of Carollian time $t_c$ together with the dynamical structure of the system. As we can see, it is easier to study the distribution of the scale dimensions and construct the dictionary using the operators in $k$ space while it may be more convenient to study the symmetries and the evolution of the system in the proposed $3d$ spacetime. We will not go into detail of the integral transform and leave the explicit form of $G(t_c,k)$ for further investigation.

The key feature for the proposed flat/celestial CFT dictionary is that it reduces two dimensions from the bulk to boundary celestial sphere. The duality relates the bulk theory to a Euclidean CFT on the sphere, with the time dependence captured by the map of a single bulk 4d field to an infinite tower of CFT operators. Many subtle questions remain about the recovery of unitary from the dual perspective. The scale dimensions of the CFT operators are complex therefore the Euclidean CFT is not unitary, yet many of the standard results used extensively in two dimensional CFTs, such as Cardy's formula, rely on unitarity. Recovery of unitarity from the dual perspective would rely on understanding how the boundary data in $k$ space can be reinterpreted in the $t_c$ domain. In particular, this would be necessary to explore how black hole information is recovered at the quantum level. 

% Through the flat/CFT correspondence, one may conclude the scattering process with gravity in the bulk is not unitary, i.e. the S matrix is not unitary, therefore the information is lost during the evaporation of the black hole \cite{Hawking:1975vcx,Hawking:1976ra}.

In our construction of the flat/CFT dictionary, one can see that the boundary correlation functions are determined by the coefficients $a^{\pm}_{lm}(k)$ which carry the information about the bulk solution. These coefficients are determined by specifying the data on the Cauchy surface of initial time and they govern the dynamical evolution of the system. To construct a proper defined quantum field theory, one should understand how constraints such as causality, Lorentz invariance and the cluster decomposition principle are related to this data. We will leave deeper exploration of such relations to further work.

We noted that one may use the data on two copies of the Euclidean AdS hyperboloid together with the equation of motion to reconstruct the linearised field in the whole Minkowski spacetime. However, one should note that these two AdS surfaces are not Cauchy surfaces according to the standard definitions \cite{Hawking:1973uf,Wald:1984rg}. A deeper understanding of the underlying structure will be helpful to study scattering amplitudes and the causal properties of spacetime.

%%%%%%%%%%%%%%%%%%%%%%%%%%%%%%%%%%%%%%%%%%%%%%%%%%%%%%%%%%%%%%%%%%%%%%%%%
%%%%%%%%%%%%%%%%%%%%%%%%%%%%%%%%%%%%%%%%%%%%%%%%%%%%%%%%%%%%%%%%%%%%%%%%%
\section*{Acknowledgments}
%%%%%%%%%%%%%%%%%%%%%%%%%%%%%%%%%%%%%%%%%%%%%%%%%%%%%%%%%%%%%%%%%%%%%%%%%
%%%%%%%%%%%%%%%%%%%%%%%%%%%%%%%%%%%%%%%%%%%%%%%%%%%%%%%%%%%%%%%%%%%%%%%%%
MT is supported in part by the Science and Technology Facilities Council (Consolidated Grant “Exploring the Limits of the Standard Model and Beyond”). ZH would like to thank his father Qinghe Hao and mother Xiulan Xu for providing funding for the tuition and accommodation fees when studying his PhD at the University of Southampton. ZH would also like to thank Federico Capone, Enrico Parisini and Kostas Skenderis for various discussions on celestial holography throughout the development of this work.
\appendix

\renewcommand{\theequation}{\Alph{section}.\arabic{equation}}

\setcounter{section}{0}

\section*{Appendix}
\section{Coordinates}\label{coordinates}
\qquad In this section, we will introduce various kinds of coordinates for Minkowski space that are convenient for us to reduce the data to the AdS hyperboloid, which are used many times in this article. The flat space time is described by the metric $\eta_{\mu\nu}$ for $\mu,\nu=0,1,2,3$, with diagonal elements $\eta_{00}=-1$ $\eta_{11}=\eta_{22}=\eta_{33}=1$, written us 
\begin{equation}
    ds^2=\eta_{\mu\nu}dX^\mu dX^\nu=-(dX^0)^2+(dX^1)^2+(dX^2)^2+(dX^3)^2,\label{metric}
\end{equation}
in which $(X^0,X^1,X^2,X^3)$ are the chosen coordinates. Here we just focus on the four dimensional spacetime and the codimension one ${\rm AdS}_3$  hypersurface characterised by the radius $\tau$ can be treated as the embedding
\begin{equation}
    -(X^0)^2+(X^1)^2+(X^2)^2+(X^3)^2=-\tau^2, \label{embedding}
\end{equation}
where we should note that here the flat Minkowski space is the physical space and the ${\rm AdS}_3$ surfaces are introduced for the decomposition of data. The timelike wedge in Minkowski which can be foliated by the AdS surfaces are so called Milne wedge.

Moreover, given such foliation, one can introduce global coordinates $(\tau,\eta,\theta,\phi)$ to cover the Milne wedge. The transformation is given by
\begin{eqnarray}
   && X^0=\tau\;\cosh{\eta},\\
   && X^1=\tau\;\sin{\theta}\;\sin{\phi}\;\sinh{\eta},\\
   && X^2=\tau\;\sin{\theta}\;\cos{\phi}\;\sinh{\eta},\\
    && X^3=\tau\;\cos{\theta}\sinh{\eta},
\end{eqnarray}
in which one can see the relation \eqref{embedding} is automatically satisfied. $(\theta, \phi)$ are coordinates on the sphere and $\tau$ is the radius of the AdS surface. The spatial distance from the origin on the hyperboloid is described by $\sinh{\eta}$. In the global coordinate, the metric now becomes 
\begin{equation}
    ds^2=-d\tau^2+ \tau^2\left(d\eta^2+\sinh^2{\eta}\;d\Omega_2^2\right),
\end{equation}
in which the metric on the standard sphere $S^2$ is given by
\begin{eqnarray}
    d\Omega_2^2&=&d\theta^2+\sin^2{\theta}d\phi^2 \\
      &=& \frac{4}{(1+z\bar{z})^2}\;dzd\bar{z}=2\gamma_{z\bar{z}}\;dzd\bar{z}.
\end{eqnarray}
The complex coordinates $(z,\bar{z})$ on the plane are obtained by the stereographic projection from the sphere
\begin{equation}
    z=e^{i\phi}\tan{\frac{\theta}{2}}\qquad \bar{z}=e^{-i\phi}\tan{\frac{\theta}{2}}.
\end{equation}
As we have mentioned, the value $\sinh{\eta}$ makes more sense as a physical quantity thus one can define $\rho=\sinh{\eta}$ then the metric now becomes
\begin{equation}
    ds^2=-d\tau^2+\tau^2\left(\frac{d\rho^2}{1+\rho^2}+2\rho^2\gamma_{z\bar{z}}dzd\bar{z}\right),
\end{equation}
which is the standard form of Milne coordinates in the literature.

To study a single AdS surface, sometimes it is more convenient to introduce Poincare coordinates $(t,x,y)$ defined as
\begin{equation}
    t=\frac{1}{X^0+X^3},\qquad x=\frac{X^1}{X^0+X^3},\qquad y=\frac{X^2}{X^0+X^3},
\end{equation}
and after setting $\tau=1$, one can pull back the metric to the AdS surface then obtain
\begin{eqnarray}
    ds^2_{\rm AdS_3}=\frac{dt^2+dx^2+dy^2}{t^2}=\frac{dt^2+d\omega d\bar{\omega}}{t^2}, \label{poin}
\end{eqnarray}
in which $\omega=x+iy$. In terms of global coordinates, the Poincar$\rm \acute{e}$ coordinates can be written as
\begin{eqnarray}
    t&=&\frac{1}{\cosh{\eta}+\cos{\theta}\sinh{\eta}},\\
    x&=&\frac{\sin{\phi} \sin{\theta}\sinh{\eta}}{\cosh{\eta}+\cos{\theta}\sinh{\eta}},\\
    y&=&\frac{\cos{\phi} \sin{\theta}\sinh{\eta}}{\cosh{\eta}+\cos{\theta}\sinh{\eta}},
\end{eqnarray}
and $(\omega,\bar{\omega})$ takes the form of
\begin{equation}
    \omega=\frac{e^{i\phi} \sin{\theta}\sinh{\eta}}{\cosh{\eta}+\cos{\theta}\sinh{\eta}},\qquad \bar{\omega}=\frac{e^{-i\phi} \sin{\theta}\sinh{\eta}}{\cosh{\eta}+\cos{\theta}\sinh{\eta}}.
\end{equation}
One should note that, in the large $\eta$ limit, $(\omega,\bar{\omega})$ will tend to $(z,\bar{z})$ thus it becomes complex coordinates of celestial sphere on the boundary.

In terms of Poincare coordinates, the boundary-bulk propagator $K_{\Delta}(t,x,y;x',y')$ for massless fields is given by
\begin{equation}
   K(t,x,y;x',y')=\frac{\Gamma(\Delta)}{\pi\Gamma(\Delta-1)}\left( \frac{t}{t^2+(x-x')^2+(y-y')^2}\right)^\Delta=\frac{N_\Delta}{\pi}\left( \frac{t}{t^2+(\omega-z')(\bar{\omega}-\bar{z}')}\right)^\Delta,
\end{equation}
in which $(x',y')$ are the points on the boundary and $z'=x'+iy'$. We set $N_\Delta=\Gamma(\Delta)/\Gamma(\Delta-1)$ for simplicity. Following the dictionary for the particles of mass $M$, we have $\Delta=1+\sqrt{1+M^2}$. The propagator could also be written in terms of global coordinates and at large $\rho$, one can check it takes the form 
\begin{equation}
    K^{\rho=\infty}(\rho,z;z')=\frac{(1+z\bar{z})^\Delta}{\pi\rho^\Delta}\;\frac{N_\Delta}{|z-z'|^{2\Delta}}=\frac{2^{\frac{\Delta}{2}}}{\pi\Omega_2(z)^{\frac{\Delta}{2}}\;\rho^\Delta}\;\frac{N_\Delta}{|z-z'|^{2\Delta}},\label{propagaotor}
\end{equation}
in which $\Omega_2(z)dzd\bar{z}=d\Omega_2$ is the volume form of the standard sphere in terms of complex coordinates. From the distribution point of view, the boundary-bulk propagators are in fact equivalent to the delta function between boundary points \cite{Witten:1998qj}, i.e, we have 
\begin{equation}
    \delta(z-z')= t^{\Delta-2} K(\rho,z;z')=\frac{2^{\Delta-1}}{\pi \rho^{2\Delta-2}\Omega_2(z)^{\Delta-1}}\frac{N_\Delta}{|z-z'|^{2\Delta}}+\cdots, \label{bulk-boundary}
\end{equation}
in which we have done the expansion of $K(\rho,z;z')$ at large radius $\rho$. In this article, we are interested in the correlation functions on the plane which is related to the sphere correlation functions by the conformal transformation and the bulk-boundary propagator is then given by
\begin{equation}
    K(\rho,z;z')=\frac{1}{\pi\rho^{\Delta}}\frac{N_\Delta}{|z-z'|^{2\Delta}}.
\end{equation}
Therefore, a generic field in the bulk with boundary behaviour $\varphi(\rho,z)\sim \rho^{\Delta-2}\varphi(z)$ can be expressed as 
\begin{equation}
    \varphi(\rho,z)=\int_{M_2}  dz'd\bar{z}' \;K(\rho,z;z')\varphi(z'), \label{field-source}
\end{equation}
in which the integral is over the two-dimension plane and we have used the relation $\delta(z-z')\sim \frac{1}{\rho^{2\Delta-2}}\frac{1}{|z-z'|^{2\Delta}}$. Here we should note that, by considering the property of the Green function
\begin{eqnarray}
    \int_{M_2} d^2z' \delta(z-z')\delta(z'-z'')=\delta(z-z'')
\end{eqnarray}
we have the contracting relation for the propagator
\begin{equation}
    \int_{M_2}dz'd\bar{z}' \frac{ N_\Delta}{\rho^{2\Delta-2}}\frac{1}{|z-z'|^{2\Delta}}\frac{1}{|z'-z''|^{2\Delta}}=\frac{\pi}{|z-z''|^{2\Delta}}, \label{contract}
\end{equation}
which turns out to be useful in simplifying the calculation.

\section{Solutions}\label{solutions}
\qquad In this section, we will present the solution of equation on the AdS hyperboloid \eqref{effective} written as
\begin{equation}
    \left(-\partial^2_\eta -2({\rm coth}\eta)\partial_\eta+l(l+1){\rm csch}^2\eta+k^2\right)\phi_{l}(\eta;k)=0, \label{radial}
\end{equation}
for $\rho=\sinh{\eta}$. Solutions can be found at the boundary and origin respectively and written as the expansion of proper basis. We should note that the basis at the origin and boundary are dependent and they are related via transformation, which we will see in the end of this section. Mode solutions for Lorentzian AdS have been studied in \cite{Balasubramanian:1998sn} while solutions for dS modes haven been studied in \cite{Liu:2021tif,laddha2022squinting}. 
\subsection*{Behaviour at the boundary}
\qquad For the solution at the boundary, we first choose to write them in terms of hypergeometric functions and then transform them into the associated Legendre functions. In order to transform the equation into the standard form for hypergeometric functions, we write the solution into the form of
\begin{equation}
    \phi_{lk}(\eta)=\frac{f_{\beta l}(\frac{1}{\sinh^2{\eta}})}{\sinh^{\beta+1}{\eta}},
\end{equation}
in which $\beta^2=1+k^2$ and $f$ depends on $\eta$ for $\eta \geq 0$. Now the equation \eqref{radial} becomes
\begin{equation}
    4x(x+1)f_{\beta l}''(x)+2(2(1+\beta)+(3+2\beta)x)f_{\beta l}'(x)-(l(l+1)-\beta(\beta+1))f_{\beta l}(x)=0,
\end{equation}
in which $x$ is defined as
\begin{equation}
        x=\frac{1}{\sinh^2{\eta}}.
\end{equation}
Here we should note that the above equation is still not in the form of hypergeometric equation because of the $x(x+1)$ term in front of $f''_{\beta l}(\eta)$. Thus we further do the transformation $x\rightarrow x-1$ then obtain the equation
\begin{equation}
    x(1-x)p_{\beta l}''(x)-\frac{1}{2}((3+2\beta)x-1)p_{\beta l}'(x)+\frac{1}{4}(l(l+1)-\beta(1+\beta))p_{\beta l}(x)=0,\label{hyper}
\end{equation}
in which $p_{\beta l}(x)$ is defined as
\begin{equation}
    p_{\beta l}(x)=f_{\beta l}(x-1).
\end{equation}
Given the equation \eqref{hyper}, one can write down the solution at $x=1$ as  
\begin{equation}
    p_{\beta l}(x)={}_2F_1(\frac{1}{2}+\frac{l}{2}+\frac{\beta}{2},-\frac{l}{2}+\frac{\beta}{2}\;;\;1+\beta\;;\;1-x)
\end{equation}
therefore the $f_{\beta l}(\eta)$ is then deduced to be
\begin{equation}
    f_{\beta l}\left(\frac{1}{\sinh^{2}{\eta}}\right)={}_2F_1\left(\frac{1}{2}+\frac{l}{2}+\frac{\beta}{2},-\frac{l}{2}+\frac{\beta}{2}\;;\;1+\beta\;;\;-\frac{1}{\sinh^2(\eta)}\right). \label{solu}
\end{equation}
Furthermore, after applying the transformation for hypergeomtric functions
\begin{equation}
    {}_2F_1(\frac{a+c-1}{2},\frac{c-a}{2};\;c\;; 4z(1-z))=(1-z)^{1-c} {}_2F_1(1-a,a\;;\;c\;;z)
\end{equation}
for 
\begin{equation}
    c=1+\beta ,\qquad a=l+1,\qquad z=\frac{1}{2}(1-\coth{\eta})
\end{equation}
to the solution \eqref{solu}, we get
\begin{equation}
    \phi_{l\beta}(\eta)=2^\beta \frac{e^{-\beta\eta}}{\sinh(\eta)}\;{}_2F_1(-l,l+1\;;\; 1+\beta\; ;\; \frac{1}{2}(1-\coth\eta)).
\end{equation}
Noting that $\beta$ could take both of the value $\beta_\pm=\pm\sqrt{1+k^2}$, one finally concludes the two independent solutions are 
\begin{equation}
    \phi_{l}(\eta;\beta_+)=\frac{\Gamma(1-\beta_+)}{(-2)^{\beta_+}}\frac{P^{\sqrt{1+k^2}}_l(\coth{\eta})}{\sinh{\eta}},\qquad \phi_{l}(\eta;\beta_-)=\frac{\Gamma(1-\beta_-)}{(-2)^{\beta_-}}\frac{P^{-\sqrt{1+k^2}}_l(\coth{\eta})}{\sinh{\eta}},
\end{equation}
in which we have taken the factor $\Gamma(1\pm\beta)$ into consideration.
\subsection*{Behaviour at the origin}
\qquad To study the behaviour of the solution at the origin, denoted as $\chi_{l}(\eta;k)$, we choose to write the function into the form of 
\begin{equation}
    \chi_{l}(\eta;k)=\sinh^a{\eta}\; f_{\beta l}(\sinh^2{\eta})
\end{equation}
in which $a$  should satisfy the relation
\begin{equation}
    a(a+1)=l(l+1)
\end{equation}
so that the equation can be recast into the hypergeometric form
\begin{equation}
    x(1-x)q''_{\beta l}(x)-\frac{1}{2}(-1+2(2+a)x)q'_{\beta l}(x)+\frac{1}{4}(\beta^2-a^2-2a)q_{\beta l}(x)=0,
\end{equation}
in which again $\beta^2$ takes the value $1+k^2$ and the function $q_{l\beta}(x)$ is defined as
\begin{equation}
    q_{\beta l}(x)=f_{\beta l}(x-1).
\end{equation}
Given the hypergeometic equation, solutions are then deduced to be
\begin{equation}
    f_{\beta l}(\sinh^2{\eta})={}_2F_1(\frac{1}{2}+\frac{a}{2}+\frac{\beta}{2},\frac{1}{2}+\frac{a}{2}-\frac{\beta}{2}\;;\; \frac{1}{2}\;;\; \cosh^2{\eta}),
\end{equation}
in which we have set $\beta=\sqrt{1+k^2}$. More precisely, for $a=l$ we have
\begin{equation}
    \chi_{ l}^1(\eta;k)=\sinh^l(\eta) \;{}_2F_1(\frac{1}{2}+\frac{l}{2}+\frac{\beta}{2},\frac{1}{2}+\frac{l}{2}-\frac{\beta}{2}\;;\; \frac{1}{2}\;;\; \cosh^2{\eta})
\end{equation}
while for $a=-1-l$ the solution becomes
\begin{equation}
    \chi_{l}^2(\eta;k)=\sinh^{-l-1}(\eta) \;{}_2F_1(-\frac{l}{2}-\frac{\beta}{2},-\frac{l}{2}+\frac{\beta}{2}\;;\; \frac{1}{2}\;;\; \cosh^2{\eta}).
\end{equation}
Here, we are just interested in the solution $\chi_{l}^l(\eta;k)$ since it is the regular solution around the origin for $l\geq 0$ and one can verify that, by using the transformation rule for hypergeometric function 
\begin{eqnarray}
    {}_2F^1(a,b;c;z)&=&\frac{(1-z)^{-a}\Gamma(c)\Gamma(b-a)}{\Gamma(b)\Gamma(b-a)} {}_2F^1\left(a,c-b;a-b+1;\frac{1}{1-z}\right)\\
    &+&(1-z)^{-b}\frac{\Gamma(c)\Gamma(a-b)}{\Gamma(a)\Gamma(c-b)}{}_2F^1\left( b,c-a; b-a+1;\frac{1}{1-z} \right),
\end{eqnarray}
it can be written in terms of the solution at the boundary as
\begin{equation}
    \chi^1_{ l}(\eta;k)=C^+_l(k)\phi_{ l}(\eta;\beta_+)+C^-_l(k)\phi_{l}(\eta;\beta_-),\label{C1}
\end{equation}
in which $C^\pm_l(k) $ are the coefficients given by
\begin{equation}
    C^+_{ l}(k)=(-i)^{1+l+\beta}\frac{\Gamma(\frac{1}{2})\Gamma(-\beta)}{\Gamma(\frac{1}{2}+\frac{l}{2}-\frac{\beta}{2})\Gamma(-\frac{l}{2}-\frac{\beta}{2})}\label{C2}
\end{equation}
and
\begin{equation}
    C^-_{ l}(k)=(-i)^{1+l-\beta}\frac{\Gamma(\frac{1}{2})\Gamma(\beta)}{\Gamma(\frac{1}{2}+\frac{l}{2}+\frac{\beta}{2})\Gamma(-\frac{l}{2}+\frac{\beta}{2})}.
\end{equation}
\subsection*{Ratio}
\qquad In order to obtain the CFT two-point function on the celestial sphere, one should calculate the functional derivative of the one-point function with respective to the source. Moreover, with the help of AdS/CFT dictionary, the functional derivative is given by the ratio of coefficients, written as
\begin{equation}
    \frac{C^-_l(k)}{C^+_l(k)}=(-1)^\beta\frac{\Gamma(\beta)\Gamma(\frac{1}{2}+\frac{l}{2}-\frac{\beta}{2})\Gamma(-\frac{l}{2}-\frac{\beta}{2})}{\Gamma(-\beta)\Gamma(\frac{1}{2}+\frac{l}{2}+\frac{\beta}{2})\Gamma(-\frac{l}{2}+\frac{\beta}{2})}.\label{ratio}
\end{equation}
To simplify above expression, we first use the recurrence formula for Gamma function given by
\begin{eqnarray}
    \Gamma(-\frac{l}{2}-\frac{\beta}{2})(-\frac{l}{2}-\frac{\beta}{2})(-\frac{l}{2}+1-\frac{\beta}{2})\cdots (\frac{l}{2}-1-\frac{\beta}{2})=\Gamma(\frac{l}{2}-\frac{\beta}{2})
\end{eqnarray}
and
\begin{eqnarray}
    \Gamma(-\frac{l}{2}+\frac{\beta}{2})(-\frac{l}{2}+\frac{\beta}{2})(-\frac{l}{2}+1+\frac{\beta}{2})\cdots (\frac{l}{2}-1+\frac{\beta}{2})=\Gamma(\frac{l}{2}+\frac{\beta}{2})
\end{eqnarray}
to transform $\Gamma(-\frac{l}{2}\pm \frac{\beta}{2})$ into $\Gamma(\frac{l}{2}\pm \frac{\beta}{2})$. Therefore, one can write the ratio \eqref{ratio} into the form of
\begin{equation}
(-1)^\beta\;\frac{\Gamma(\beta)\Gamma(\frac{1}{2}+\frac{l}{2}-\frac{\beta}{2})\Gamma(\frac{l}{2}-\frac{\beta}{2})}{\Gamma(-\beta)\Gamma(\frac{1}{2}+\frac{l}{2}+\frac{\beta}{2})\Gamma(\frac{l}{2}+\frac{\beta}{2})}\times \frac{(-\frac{l}{2}+\frac{\beta}{2})(-\frac{l}{2}+1+\frac{\beta}{2})\cdots (\frac{l}{2}-1+\frac{\beta}{2})}{(-\frac{l}{2}-\frac{\beta}{2})(-\frac{l}{2}+1-\frac{\beta}{2})\cdots (\frac{l}{2}-1-\frac{\beta}{2})}.\label{ratio2}
\end{equation}
After applying the Legendre duplication formula
\begin{equation}
    \Gamma(z)\Gamma(z+\frac{1}{2})=2^{1-2z}\sqrt{\pi}\Gamma(2z)
\end{equation}
for $z=\frac{l}{2}\pm \frac{\beta}{2}$, we have 
\begin{equation}
    \Gamma(\frac{1}{2}+\frac{l}{2}\pm \frac{\beta}{2})\Gamma(\frac{l}{2}\pm \frac{\beta}{2})=\sqrt{\pi}2^{1-(l\pm \beta)}\Gamma(l\pm \beta).\label{s1}
\end{equation}
For the part on right of \eqref{ratio2}, one should notice that
\begin{equation}
\frac{(-\frac{l}{2}+\frac{\beta}{2})(-\frac{l}{2}+1+\frac{\beta}{2})\cdots (\frac{l}{2}-1+\frac{\beta}{2})}{(-\frac{l}{2}-\frac{\beta}{2})(-\frac{l}{2}+1-\frac{\beta}{2})\cdots (\frac{l}{2}-1-\frac{\beta}{2})}=(-1)^l\frac{(-\frac{l}{2}+1-\frac{\beta}{2})(-\frac{l}{2}+2-\frac{\beta}{2})\cdots (\frac{l}{2}-\frac{\beta}{2})}{(-\frac{l}{2}-\frac{\beta}{2})(-\frac{l}{2}+1-\frac{\beta}{2})\cdots (\frac{\beta}{2}-1-\frac{\beta}{2})}=(-1)^l\frac{\beta-l}{\beta+l}.\label{s2}
\end{equation}
After substituting \eqref{s1} and \eqref{s2} into \eqref{ratio2}, one has 
\begin{eqnarray}
    \frac{C_l^-(k)}{C_l^+(k)}&=&(-1)^{l+\beta}\;2^{2\beta}\; \frac{\Gamma(\beta)\Gamma(l-\beta)(\beta-l)}{\Gamma(-\beta)\Gamma(l+\beta)(\beta+l)}\\
    &=& (-1)^{l+\beta+1}\; 4^\beta\; \frac{\Gamma(\beta)\Gamma(l-\beta+1)}{\Gamma(-\beta)\Gamma(l+\beta+1)}\\
    &=& (-1)^{l+\beta+1}\; 4^\beta \; \frac{B(\beta,l-\beta+1)}{B(-\beta,l+\beta+1)},
\end{eqnarray}
where we have written the result in terms of Beta function $B(x,y)=\frac{\Gamma(x)\Gamma(y)}{\Gamma(x+y)}$ in the third line.
\section{Harmonic Modes}\label{Harmonic}
\qquad In this section we will present an alternative way to calculate the two-point correlation functions different from the bulk-boundary approach used in section \ref{Dictionary}. The strategy here is that one can study the function on the sphere in terms of the discrete harmonic variables rather than the continuous complex coordinates and the functional variation in complex coordinates becomes ratio between functions in the discrete coordinates. First we will review the calculation in the context of AdS/CFT then generalise this into flat/CFT and at the end argue the equivalence between the propagator approach and the approach for doing the ratio. Such method allows us to calculate the two point function for asymptotic flat space when finding the solution of the propagator becomes hard.

Like the Fourier transform between the spacetime and momentum, the transformation between the discrete mode variables $(l,m)$ and complex coordinates $(z,\bar{z})$ on the sphere  
\begin{equation}
    (l,m)\longleftrightarrow (z,\bar{z})
\end{equation}
are related by the spherical harmonics $Y^l_m(z,\bar{z})$. More precisely, as shown in \eqref{tran}, the transformation is realised via the expansion 
\begin{equation}
    F_k(\rho,z,\bar{z}):=\sum_{lm}F_{k,l,m}(\rho,z,\bar{z})= \sum_{l,m}\phi_l(\rho;k)Y_m^l(z,\bar{z})
\end{equation}
in which $F_k(\rho,z,\bar{z})$ is the spatial $k$ mode that depends on $(\rho,z,\bar{z})$ and $\phi_l(\rho;k)$ is the associated expression in the mode variables $(\rho,l,m)$. The $m$ dependence is suppressed since the equation of motion on the AdS hyperboloid does not depend on $m$ \footnote{In fact, it is more appropriate to use the notion $\phi_{lm}(\rho;k)$ here even though the solution dose not depend on $m$ explicitly.}. Given the solution $\phi_l(\rho;\beta_\pm)$ and their asymptotic expansion at infinity
\begin{equation}
    \phi_l(\rho;\beta_\pm)=\rho^{\beta_\pm-1}(\phi^\pm_l(k)+\mathcal{O}(\frac{1}{\rho^2})), \label{exp}
\end{equation}
one can immediately obtain the dictionary for AdS/CFT in the form of mode variables $(m,l)$, written as
\begin{equation}
 \hat{\mathcal{J}}_{lm}(k)=\phi_l^+(k)  \qquad \langle \hat{\mathcal{O}}_{lm}(k)\rangle=-2i\beta_+\; \phi^-_l(k) \qquad {\rm for } \qquad  -l\leq m \leq l,
\end{equation}
in which $\hat{\mathcal{J}}_{lm}(k)$ and $\langle \hat{\mathcal{O}}_{lm}(k)\rangle$ are the corresponding source and one-point function that lives on the boundary celestial sphere. Here they are not required to be physical operators and sources thus we can treat them as virtual particles by construction. In terms of $(z,\bar{z})$ coordinates, they should have the form of
\begin{equation}
    \hat{\mathcal{J}}(z,\bar{z};k)=\sum_{l,m}\phi_l^+(k)Y_m^l(z,\bar{z}),\qquad \langle \hat{\mathcal{O}}(z,\bar{z};k)\rangle=-\sum_{lm}2i\beta_+\; \phi^-_l(k)Y_m^l(z,\bar{z}). 
\end{equation}

Here we should note that $\phi_l(\rho;\beta_\pm)$ are two independent solutions at the boundary and they are singular at the origin. The regular solution can be obtained by directly solving the equation at the origin so called $\chi^1(\eta;k)$. They are solutions of the same equation at different singular points so $\phi_l(\rho;\beta_\pm)$, $\chi^1(\eta;k)$ are not independent. The transformation between them are given by
\begin{equation}
    \chi^1_{ l}(\eta;k)=C^+_l(k)\phi_{ l}(\eta;\beta_+)+C^-_l(k)\phi_{l}(\eta;\beta_-),
\end{equation}
in which we have chosen solutions at the boundary as the basis and $C^\pm_l(k)$ are coefficients determined in \eqref{C1} and \eqref{C2}. Given the above relation, one can get the functional derivative between the source and one-point function thus higher point functions can be determined. For the two-point function, we have
\begin{equation}
    \langle \hat{\mathcal{O}}_{lm}(k)\; \hat{\mathcal{O}}_{l',m'}(k) \rangle = \frac{\delta\langle \hat{\mathcal{O}}_{lm}(k)\rangle_{J}}{\delta \hat{\mathcal{J}}_{l'm'}(k)} \bigg|_{J=0}=-\delta^l_{l'}\delta^m_{m'}\;2i\beta_+\frac{C^-_l(k)}{C^+_l(k)}, \label{2p}
\end{equation}
in which the two-point function is written in terms of mode variables $(l,m)$. Here, we should note the value of $\langle \hat{\mathcal{O}}\rangle_J$ is scheme dependent and we assume that proper regularization procedure in the mode space $(l,m)$ exists so that \eqref{2p} is true for two-point function, like what has been done in momentum space \cite{Freedman:1991tk,Freedman:1998tz,skenderis2002lecture}.  To go back to the complex coordinates on the celestial sphere, one can do the sum over spherical harmonics $Y_m^l(z,\bar{z})$ then obtain
\begin{equation}
    \langle \hat{\mathcal{O}}(z,\bar{z};k)\;\hat{\mathcal{O}}(z',\bar{z}';k) \rangle=-2i\beta_+\sum_{lm} \frac{C^-_l(k)}{C^+_l(k)} Y^l_m(z,\bar{z})Y^l_{m}(z',\bar{z}'). \label{trans}
\end{equation}
Here we should note that the two-point function on the sphere is obtained by summing over two discrete variables $(l,m)$ while one can also just do the sum over variable $m$  and obtain the $l$-mode source, one-point function
\begin{equation}
    \hat{\mathcal{J}}_l(z,\bar{z};k)=\sum _m\phi_l^+(k) Y_m^l(z,\bar{z}) \qquad \langle \hat{\mathcal{O}}_l(z,\bar{z};k)\rangle=-2i\beta_+\sum _m\phi_l^-(k) Y_m^l(z,\bar{z}), \label{lmode}
\end{equation}
and the corresponding two-point function is given by
\begin{equation}
    \langle \hat{\mathcal{O}}_l(z,\bar{z};k)\;\hat{\mathcal{O}}_l(z',\bar{z}';k) \rangle=-2i\beta_+\sum_{m} \frac{C^-_l(k)}{C^+_l(k)} Y^l_m(z,\bar{z})Y^l_{m}(z',\bar{z}').\label{l-2pt}
\end{equation}
\subsubsection*{Two-point Function}
\qquad To study the dictionary for flat space in a more precise way, we consider a generic $k$ mode $f(\tau,\rho,z,\bar{z};k)$ for on-shell field $\Phi(\tau,\rho,z,\bar{z})$ defined in \eqref{kmode}, or equivalently 
\begin{equation}
 f(\tau,\rho,z,\bar{z};k)=\sum_{lm}\int d\omega f_{w,k,l,m}(\tau,\rho,z,\bar{z})\Tilde{\Phi}(w,k,l,m).
\end{equation}
Following the notion in \eqref{spatialde}, we choose to decompose the $k$ mode into the spatial modes therefore $f(\tau,\rho,z,\bar{z};k)$ now takes the form 
\begin{eqnarray}
f(\tau,\rho,z,\bar{z};k)&=& \sum _{lm}\bar{\Phi}(\tau,k,l,m)\phi_l(\rho;k)Y_m^l(z,\bar{z})\\
&=&\sum_{lm}(a^+_{lm}(k) f_+(\tau,k)+a^-_{lm}(k) f_-(\tau,k))\;(\phi_l(\rho;\beta_+) + \phi_l(\rho;\beta_-)) Y_m^l(z,\bar{z})   \\
&=&  f_+(\tau,k)\phi(\rho,z,\bar{z};\beta_+) +f_-(\tau,k)\tilde{\phi}(\rho,z,\bar{z};\beta_+)\\&&
  f_-(\tau,k)\phi(\rho,z,\bar{z};\beta_-)+f_-(\tau,k)\tilde{\phi}(\rho,z,\bar{z};\beta_-),
\end{eqnarray}
in which $\bar{\Phi}(\tau,k,l,m)$ and $\phi_l(\rho,k)$ are modes that depend on $\tau$ and $\rho$. In the first line, we have summed over the two discrete variables $l,m$ and also made the $\tau$-mode $l,m$ dependent by introducing the coefficients $a_{lm}^{\pm}(k)$ \footnote{Here, we should note that coefficients $a_{lm}^{\pm}(k)$ play the same role as $\psi_\pm(p)$ in \eqref{co1} or $\psi(p)$ in \eqref{co2} for fixed $l,m$. }. They are determined by the initial data. In the third line, we rearrange them into the $\tau$ mode functions and highlight their asymptotic behaviour according to $\beta_\pm$. The function $\phi(\rho,z,\bar{z},\beta_\pm)$, $\tilde{\phi}(\rho,z,\bar{z},\beta_\pm)$  are given by
\begin{eqnarray}
    \phi(\rho,z,\bar{z};\beta_\pm)&=&\sum_{lm} a^+_{lm}(k)\;\phi_l(\rho;\beta_\pm) Y_m^l(z,\bar{z}) \\
      \tilde{\phi}(\rho,z,\bar{z};\beta_\pm)&=&\sum_{lm} a^-_{lm}(k)\;\phi_l(\rho;\beta_\pm) Y_m^l(z,\bar{z}).
\end{eqnarray}
Moreover, using the asymptotic expansion \eqref{exp} for $\phi_l(\rho;\beta_\pm)$
we obtain the leading contribution for $\phi(\rho,z,\bar{z};\beta_\pm)$ and $\tilde{\phi}(\rho,z,\bar{z};\beta_\pm)$ written as
\begin{eqnarray}
    \phi^\pm(z,\bar{z};k)&=&\sum_{lm} a^+_{lm}(k)\;\phi^\pm_l(k) Y_m^l(z,\bar{z}) \\
      \tilde{\phi}^\pm(z,\bar{z};k)&=&\sum_{lm} a^-_{lm}(k)\;\phi_l^\pm(k) Y_m^l(z,\bar{z}).
\end{eqnarray}
Now, given the above asymptotic expansion, we rewrite the flat/CFT dictionary \eqref{vev1} into
\begin{equation}
 \mathcal{J}(z,\bar{z};k)=\sum _{lm}a_{lm}^-(k)\mathcal{J}_{lm}(k) Y_l^m(z,\bar{z}) \qquad \langle \mathcal{O}(z,\bar{z};k)\rangle=\sum _{lm} a^+_{lm}(k)\langle\mathcal{O}_{lm}(k)\rangle Y_m^l(z,\bar{z}),
 \end{equation}
 \begin{equation}
 \tilde{\mathcal{J}}(z,\bar{z};k)=\sum _{lm}a_{lm}^+(k)\hat{\mathcal{J}}_{lm}(k) Y_l^m(z,\bar{z}) \qquad \langle \tilde{\mathcal{O}}(z,\bar{z};k)\rangle=\sum _{lm} a^-_{lm}(k)\langle\hat{\mathcal{O}}_{lm}(k)\rangle Y_m^l(z,\bar{z}),
 \end{equation}
 from which we can see there is a pair of source and one-point function $\{ \mathcal{J},\mathcal{O} \}$, $\{ \tilde{\mathcal{J}},\tilde{\mathcal{O}} \}$ and they are combination of the source and one-point functions introduced in the AdS/CFT dictionary. Here we should note that the source and one-point functions $\{ \mathcal{J},\mathcal{O} \}$, $\{\tilde{\mathcal{J}}, \tilde{\mathcal{O}}\}$ now become physical and their existence does not rely on the AdS/CFT dictionary i.e., one could study them without writing them in terms of AdS modes $\{\hat{\mathcal{J}}_{lm},\hat{\mathcal{O}}_{lm}\}$. Given the above dictionary, one can deduce the two-point function
\begin{equation}
      \langle\tilde{\mathcal{O}}(z,\bar{z};k)\tilde{\mathcal{O}}(z',\bar{z}';k)\rangle=\frac{1}{N_k}\sum_{lm}\frac{a^-_{lm}(k)}{a^+_{lm}(k)}\langle\hat{\mathcal{O}}_{lm}(k)\hat{\mathcal{O}}_{l,m}(k)\rangle Y^l_{m}(z,\bar{z})Y^l_{m}(z',\bar{z}'),\label{2pth1} 
 \end{equation}
\begin{equation}
      \langle\mathcal{O}(z,\bar{z};k)\mathcal{O}(z',\bar{z}';k)\rangle=\frac{1}{N_k}\sum_{lm}\frac{a^+_{lm}(k)}{a^-_{lm}(k)}\langle\hat{\mathcal{O}}_{lm}(k)\hat{\mathcal{O}}_{l,m}(k)\rangle Y^l_m(z,\bar{z})Y^l_{m}(z',\bar{z}'),\label{2pth2}
 \end{equation}
in which the $(l.m)$ mode two-point functions are given in\eqref{2p}. Here we should note that $a^-_{lm}/a^+_{lm}=a^+_{lm}/a^-_{lm}=0$ if $a^-_{lm}=0$ or $a^+_{lm}=0$.

During the calculation, we assume the coefficients $a_{lm}(k)$ determined by the initial data are $(l,m)$ dependent. In fact, we can simplify the coefficients if there is a rotating symmetry for the solution on the sphere thus the coefficients will be $m$ independent and we label them as $a_l(k)$. In this case, the $k$ mode will be written as 
\begin{eqnarray}
f(\tau,\rho,z,\bar{z};k)
&=&\sum_{lm}(a^+_{l}(k) f_+(\tau,k)+a^-_{l}(k) f_-(\tau,k))\;(\phi_l(\rho;\beta_+) + \phi_l(\rho;\beta_-)) Y_m^l(z,\bar{z}).
\end{eqnarray}
The flat/CFT dictionary remains the same while the source and one-point function will be written in terms of the shorter form 
\begin{equation}
 \mathcal{J}(z,\bar{z};k)=\sum _la_l^-(k)\hat{\mathcal{J}}_l(z,\bar{z};k) \qquad \langle \mathcal{O}(z,\bar{z};k)\rangle=\sum _l a^+_l(k)\langle\hat{\mathcal{O}}_l(z,\bar{z};k)\rangle,
 \end{equation}
\begin{equation}
 \tilde{\mathcal{J}}(z,\bar{z};k)=\sum _la_l^+(k)\hat{\mathcal{J}}_l(z,\bar{z};k) \qquad \langle \tilde{\mathcal{O}}(z,\bar{z};k)\rangle=\sum _l a^-_l(k)\langle\hat{\mathcal{O}}_l(z,\bar{z};k)\rangle,
 \end{equation}
in which $\{ \mathcal{J}_l ,\mathcal{O}_l\}$ are the $l$ mode source and one-point function defined in \eqref{lmode}. As for the two-point function, following the standard functional derivative procedure, we have
 \begin{equation}
      \langle \tilde{\mathcal{O}}(z,\bar{z};k)\tilde{\mathcal{O}}(z',\bar{z}';k)\rangle=\frac{\delta \langle \tilde{\mathcal{O}}(z,\bar{z};k)\rangle_J}{\delta \tilde{\mathcal{J}}_k(z',\bar{z}';k)}\bigg|_{J=0}=\frac{1}{N_k}\sum_l\frac{a^-_l(k)}{a^+_l(k)}\langle\hat{\mathcal{O}}_l(z,\bar{z};k)\hat{\mathcal{O}}_l(z',\bar{z}';k)\rangle,
 \end{equation}
  \begin{equation}
      \langle\mathcal{O}(z,\bar{z};k)\mathcal{O}(z',\bar{z}';k)\rangle=\frac{\delta \langle\mathcal{O}(z,\bar{z};k)\rangle_J}{\delta \mathcal{J}(z',\bar{z}';k)}\bigg|_{J=0}=\frac{1}{N_k}\sum_l\frac{a^+_l(k)}{a^-_l(k)}\langle\hat{\mathcal{O}}_l(z,\bar{z};k)\hat{\mathcal{O}}_l(z',\bar{z}';k)\rangle,
 \end{equation}
 in which the $l$-mode two-point function on the right hand side are given by \eqref{l-2pt}. From the above discussion, one can see that it is not possible to simplify the coefficients $a_l(k)$ further and make them $l$ independent otherwise the pair of source and one-point function will become linearly dependent and be reduced to one copy.
\subsubsection*{Boundary-Bulk Propagator}
\qquad Following the study of the mode expansion of the fields, we know that the source of the fields can be expanded by the spherical harmonics on the sphere with coefficients $a^{\pm}_{lm}(k)$, written as
\begin{eqnarray}
    &&\mathcal{J}({z,\bar{z};k})=\tilde{\phi}^+(z,\bar{z};k)=\sum_{lm}a^-_{lm}(k)Y_m^l(z,\bar{z}),\\
    &&\tilde{\mathcal{J}}({z,\bar{z};k})=\phi^+(z,\bar{z};k)=\sum_{lm}a^+_{lm}(k)Y_m^l(z,\bar{z}).
\end{eqnarray}
Therefore, with the help of the bulk-boundary propagator $K(\rho,z;z')$, one can then write the spatial mode $\phi(\rho,z,\bar{z};k)$ and $\phi(\rho,z,\bar{z};k)$ into the form of 
\begin{eqnarray*}
 &&  \phi(\rho,z,\bar{z};k)=\int dz'd\bar{z}' K(\rho,z;z') \phi^+(z',\bar{z}';k)=\sum_{lm}a^+_{lm}(k)\int dz'd\bar{z}' K(\rho,z;z')Y^l_m(z',\bar{z}'),\\
 &&  \tilde{\phi}(\rho,z,\bar{z};k)=\int dz'd\bar{z}' K(\rho,z;z') \tilde{\phi}^+(z',\bar{z}';k)=\sum_{lm}a^-_{lm}(k)\int dz'd\bar{z}' K(\rho,z;z')Y^l_m(z',\bar{z}').
\end{eqnarray*}
Given such expression, together with the flat/CFT dictionary, the one-point functions are now deduced to be 
\begin{eqnarray}
    &&\mathcal{O}({z,\bar{z};k})=-2i\beta_+\phi^-(z,\bar{z};k)=-\frac{2i\beta_+}{\pi}\sum_{lm}a^+_{lm}(k)\int dz'd\bar{z}'\frac{1}{|z-z'|^{2\Delta_k}}Y_m^l(z',\bar{z}'),\\
    &&\tilde{\mathcal{O}}({z,\bar{z};k})=-2i\beta_+\tilde{\phi}^-(z,\bar{z};k)=-\frac{2i\beta_+}{\pi}\sum_{lm}a^-_{lm}(k)\int dz'd\bar{z}'\frac{1}{|z-z'|^{2\Delta_k}}Y_m^l(z',\bar{z}').
\end{eqnarray}
Moreover, by doing the functional variation with respect to the source $\mathcal{J}(z,\bar{z};k)$ and $\tilde{\mathcal{J}}(z,\bar{z};k)$, one should be able to obtain the two point functions. The functional variation between the one-point function and the source can be transformed into variation between spherical harmonics since both of the operator $\mathcal{O}$ and the source $\mathcal{J}$ are now written in terms of harmonic function $Y^l_m$. At first, as a kind of approximation, we assume that the boundary-bulk propagator is a function that do not depend on the spherical harmonics, then the two-point functions become
\begin{eqnarray}
   && \langle \mathcal{O}(z,\bar{z};k) \mathcal{O}(z',\bar{z}';k)\rangle=\frac{1}{N_k}\sum_{l\neq 0 ,m}\frac{a^+_{lm}(k)}{a^-_{lm}(k)}\frac{c_k}{|z-z'|^{2\Delta_k}},\nonumber\\
   &&\langle \tilde{ \mathcal{O}}(z,\bar{z};k) \tilde{\mathcal{O}}(z',\bar{z}';k)\rangle=\frac{1}{N_k}\sum_{l\neq 0 ,m}\frac{a^-_{lm}(k)}{a^+_{lm}(k)}\frac{c_k}{|z-z'|^{2\Delta_k}},\label{flat 2pt}
\end{eqnarray}
in which $c_k=-2i\beta_+/\pi$ is the renormalised factor. Now, we will determine the functional variation in a more precise way by decomposing the bulk-boundary into the harmonics modes
\begin{equation}
    K(\rho,z;z')=\sum_{lm}K_{lm}(\rho)Y^l_m(z,\bar{z})Y^l_{m}(z',\bar{z}')
\end{equation}
in which the function $K(\rho)$ can be treated as the coefficients. Given such decomposition, the one-point function can be written into the form 
\begin{eqnarray}
    &&\mathcal{O}({z,\bar{z};k})=-2i\beta_+\sum_{lm}a^+_{lm}(k)\int dz'd\bar{z}'K_{lm}Y^l_m(z,\bar{z})\left(Y_{m}^l(z',\bar{z}')\right)^2,\\
    &&\tilde{\mathcal{O}}({z,\bar{z};k})=-2i\beta_+\sum_{lm}a^-_{lm}(k)\int dz'd\bar{z}'K_{lm}Y^l_m(z,\bar{z})\left(Y_{m}^l(z',\bar{z}')\right)^2,
\end{eqnarray}
in which $K_{lm}=\rho^\Delta K_{lm}(\rho)$. Therefore, by calculating the functional variation with respect to the spherical harmonics $Y^l_m$, one then obtain the two point function
\begin{eqnarray}
   && \langle \mathcal{O}(z,\bar{z};k) \mathcal{O}(z',\bar{z}';k)\rangle=-\frac{4i\beta_+}{N_k}\sum_{lm}\frac{a^+_{lm}(k)}{a^-_{lm}(k)}K_{lm}Y^l_m(z,\bar{z})Y^l_{m}(z',\bar{z}')\\
   &&\langle \tilde{ \mathcal{O}}(z,\bar{z};k) \tilde{\mathcal{O}}(z',\bar{z}';k)\rangle=-\frac{4i\beta_+}{N_k}\sum_{lm}\frac{a^-_{lm}(k)}{a^+_{lm}(k)}K_{lm}Y^l_m(z,\bar{z})Y^l_{m}(z',\bar{z}').
\end{eqnarray}
One can check such expression is equivalent to the result obtained from the mode analysis calculation by making 
\begin{equation}
   4i\beta_+ K_{lm}\equiv \delta^l_{l'}\delta^m_{m'}\langle \hat{\mathcal{O}}_{lm}(k)\hat{\mathcal{O}}_{l'm'}(k)\rangle,
\end{equation}
or equivalently we have
\begin{equation}
    2K_{lm}=\frac{C^-_l(k)}{C_l^+(k)}.
\end{equation}

\bibliographystyle{JHEP}
\bibliography{ref}

\end{document}